\documentclass[]{aa} 

\usepackage[varg]{txfonts}
\usepackage{natbib,graphicx,amssymb}
\usepackage[english]{babel}
\usepackage{xspace}
\usepackage{lscape}
\usepackage{breqn}
\usepackage{url}
\usepackage{float}
\usepackage{amsmath}

\newcommand{\mum}{$\mu$m\xspace}
\newcommand{\cii}{[C\,{\sc ii}]\xspace}
\newcommand{\nii}{[N\,{\sc ii}]\xspace}
\newcommand{\oi}{[O\,{\sc i}]\xspace}
\newcommand{\oiii}{[O\,{\sc iii}]\xspace}
\newcommand{\hers}{\textit{Herschel}\xspace}
\newcommand{\spit}{\textit{Spitzer}\xspace}
\newcommand{\siii}{[S\,{\sc iii}]\xspace}
\newcommand{\siv}{[S\,{\sc iv}]\xspace}
\newcommand{\neii}{[Ne\,{\sc ii}]\xspace}
\newcommand{\neiii}{[Ne\,{\sc iii}]\xspace}
\newcommand{\hii}{H\,{\sc ii}\xspace}
\newcommand{\hi}{H\,{\sc i}\xspace}

\newcommand{\arcs}{$^{\prime\prime}$\xspace}

\usepackage{color}

\begin{document}

\title{Modeling the physical properties in the ISM \\
of the low-metallicity galaxy NGC\,4214\thanks{based on observations performed with \spit and \hers.}}

\author{
  A. Dimaratos\inst{1}
  \and D.~Cormier\inst{1}
  \and F.~Bigiel\inst{1}
  \and S.~C.~Madden\inst{2}
}

\institute{ 
Institut f\"ur theoretische Astrophysik, 
Zentrum f\"ur Astronomie der Universit\"at Heidelberg, 
Albert-Ueberle Str. 2, 69120 Heidelberg, Germany. 
\and
Laboratoire AIM, CEA/DSM - CNRS - Universit\'e Paris
  Diderot, Irfu/Service d'Astrophysique, CEA Saclay, 91191
  Gif-sur-Yvette, France 
}


\abstract{We present a model for the interstellar medium of NGC\,4214 with the objective to probe the physical conditions in the two main star-forming regions and their connection with the star formation activity of the galaxy. We used the spectral synthesis code \textsc{Cloudy} to model an \hii region and the associated photodissociation region (PDR) to reproduce the emission of mid- and far-infrared fine-structure cooling lines from the \spit and \hers space telescopes for these two regions. Input parameters of the model, such as elemental abundances and star formation history, are guided by earlier studies of the galaxy, and we investigated the effect of the mode in which star formation takes place (bursty or continuous) on the line emission. Furthermore, we tested the effect of adding pressure support with magnetic fields and turbulence on the line predictions. We find that this model can satisfactorily predict (within a factor of $\sim$2) all observed lines that originate from the ionized medium (\siv10.5\mum, \neiii15.6\mum, \siii18.7\mum, \siii33.5\mum, and \oiii88\mum), with the exception of \neii12.8\mum and \nii122\mum, which may arise from a lower ionization medium. In the PDR, the \oi63\mum, \oi145\mum, and \cii157\mum lines are matched within a factor of $\sim$5 and work better when weak pressure support is added to the thermal pressure or when the PDR clouds are placed farther away from the \hii regions and have covering factors lower than unity. Our models of the \hii region agree with different evolutionary stages found in previous studies, with a more evolved, diffuse central region, and a younger, more compact southern region. However, the local PDR conditions are averaged out on the 175\,pc scales that we probe and do not reflect differences observed in the star formation properties of the two regions. Their increased porosity stands out as an intrinsic characteristic of the low-metallicity ISM, with the PDR covering factor tracing the evolution of the regions.}

\titlerunning{Modeling the ISM of NGC\,4214}
\authorrunning{Dimaratos et al.}
\maketitle

\section{Introduction}
\label{introduction}
The physical state and structure of the interstellar medium (ISM) are important parameters for understanding the star formation in a galaxy. In a typical star-forming region, young massive stars are born and start to illuminate their parental cloud. UV photons ionize the surrounding medium, creating \hii regions, while the transition to the neutral atomic or molecular phase occurs at higher visual extinction, where the material is more effectively shielded. Far-UV (FUV) photons control the chemical activity in these regions, namely the photodissociation region (PDR; \citealt{tielens-1985}). By studying the latter, we can investigate the conditions of the molecular clouds, which in turn will be potential sites for the next episode of star formation. 

How does the propagation of radiation and the ISM composition affect
ISM observables in low-metallicity galaxies? Addressing this question is important to understand the evolution of low-metallicity galaxies, which undergo more bursty star formation than normal galaxies. Nearby star-forming dwarf galaxies present distinct
observational signatures compared to well-studied disk galaxies. Dwarfs are usually metal poor, \hi\ rich, and molecule poor as
a result of large-scale photodissociation \citep[e.g.,][]{kunth-2000,hunter-2012,schruba-2012}. Mid-IR (MIR) and far-IR (FIR) observations have revealed bright atomic lines from \hii regions (\siii, \neiii, \neii, \oiii, etc.) and PDRs (\cii, \oi) \citep[e.g.,][]{hunter-2001,madden-2006,wu-2008,hunt-2010,cormier-2015}. Their spectral energy distributions (SEDs) are also different from spiral and elliptical galaxies and indicative of altered dust properties, with a relatively low abundance of polycyclic aromatic hydrocarbons (PAHs) and perhaps a different dust composition \citep[e.g.,][]{madden-2006,galliano-2008,remy-2013}. 
It is still unknown, however, whether these differences between dwarf and disk galaxies are the direct result of recent star formation activity shaping the ISM or instead a consequence of the low-metallicity ISM that is independent of star formation activity. To answer this, one needs to observe tracers of the interplay between the ISM and various stages of star formation activity. While there are now a number of important studies available on PDR properties modeling FIR lines on large scales in various extragalactic environments \citep[e.g.,][]{kaufman-2006,vasta-2010,gracia-carpio-2011,cormier-2012,parkin-2013} or in our Galaxy under solar-metallicity conditions \citep[e.g.,][]{cubick-2008,bernard-salas-2012,bernard-salas-2015}, only a few studies are published on individual extragalactic regions \citep{mookerjea-2011,lebouteiller-2012}. Of particular interest are dwarf galaxies, where the effect due to radiative feedback is expected to be most significant. 
The goal of this paper is to investigate how the low-metallicity ISM reacts under the effects of star formation in regions that have undergone different histories. The nearby low-metallicity galaxy NGC\,4214 provides an excellent environment to perform this experiment because it has well-separated star-forming centers, one hosting a super star cluster, which allows us to study the effects of extreme star-forming conditions on the surrounding ISM. 

\begin{figure}
\centering
\includegraphics[scale=0.52]{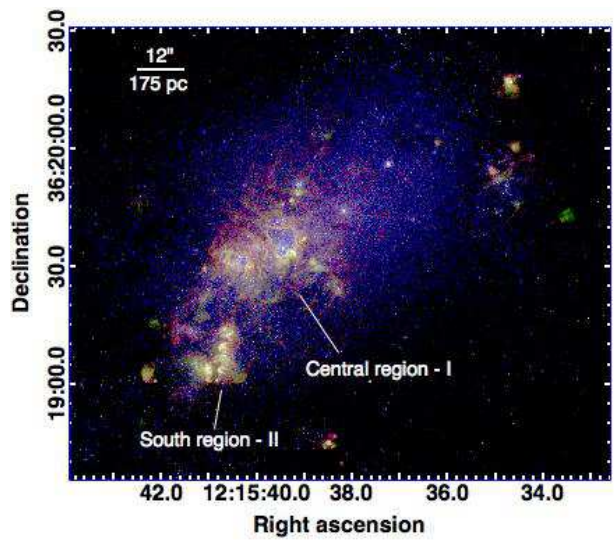}
\centering
\caption{Three-color image of NGC\,4214 using the HST WFC3 filters F438W (B, blue), F502N (\oiii, green), and F657N (H$\alpha$+\nii, red), downloaded from the Hubble Legacy Archive (\protect\url{http://hla.stsci.edu/}).}
\label{3color}
\end{figure}

NGC\,4214 is a nearby irregular galaxy located 3\,Mpc away \citep{dalcanton-2009} with a metallicity of $\sim$0.3\,Z$_{\odot}$ \citep{kobulnicky-1996} and a wealth of ancillary data. It shows various morphological characteristics such as \hi holes and shells and a spiral pattern \citep{mcintyre-1998}. 
NGC\,4214 is known to host two main, well-defined star-forming regions with recent activity (Fig.\,\ref{3color}). The largest of the two regions is found in the center of the galaxy (also referred to as NW or region~I) and contains several clusters, including a super star cluster, while the second region is found to the southeast (also referred to as SE or region~II) and is  younger and more compact. 
Using near-IR, optical, and UV data, several studies have constrained the ages of the clusters in the two main regions, which show evidence for recent star formation \citep{ubeda-2007,sollima-2013,sollima-2014}. \cite{schruba-2012} have measured the ongoing SFR of NGC\,4214 to be 0.12\,M$_{\odot}$\,yr$^{-1}$. 
The galaxy seems to have maintained its star formation in the past 10\,Gyr at an average rate of $\sim$0.02\,M$_{\odot}$\,yr$^{-1}$, with a prolonged star formation episode that occurred about 3\,Gyr ago and several shorter bursty events within the past Gyr at a rate of 0.05-0.12\,M$_{\odot}$\,yr$^{-1}$ \citep{mcquinn-2010,williams-2011}. \par

In this paper, we present observations of MIR and FIR fine-structure cooling lines in NGC\,4214, which provide key diagnostics of the physical conditions of the ISM. We focus our analysis on the two main star-forming complexes. The line emission is analyzed with radiative transfer models to characterize the ISM conditions. We take into account directly observed star formation histories and explore how they affect the IR line emission. Photometry is used for the energy budget of the models. The structure of this paper is the following: Sect.~\ref{data} describes the data, Sect.~\ref{method} describes the model, and the results are presented in Sect.~\ref{results}. We summarize and discuss our results in Sect.~\ref{discussion}.

\section{Data}
\label{data}
\subsection{{\it Herschel} data}
\begin{figure*}
\centering
\includegraphics[width=5.585cm]{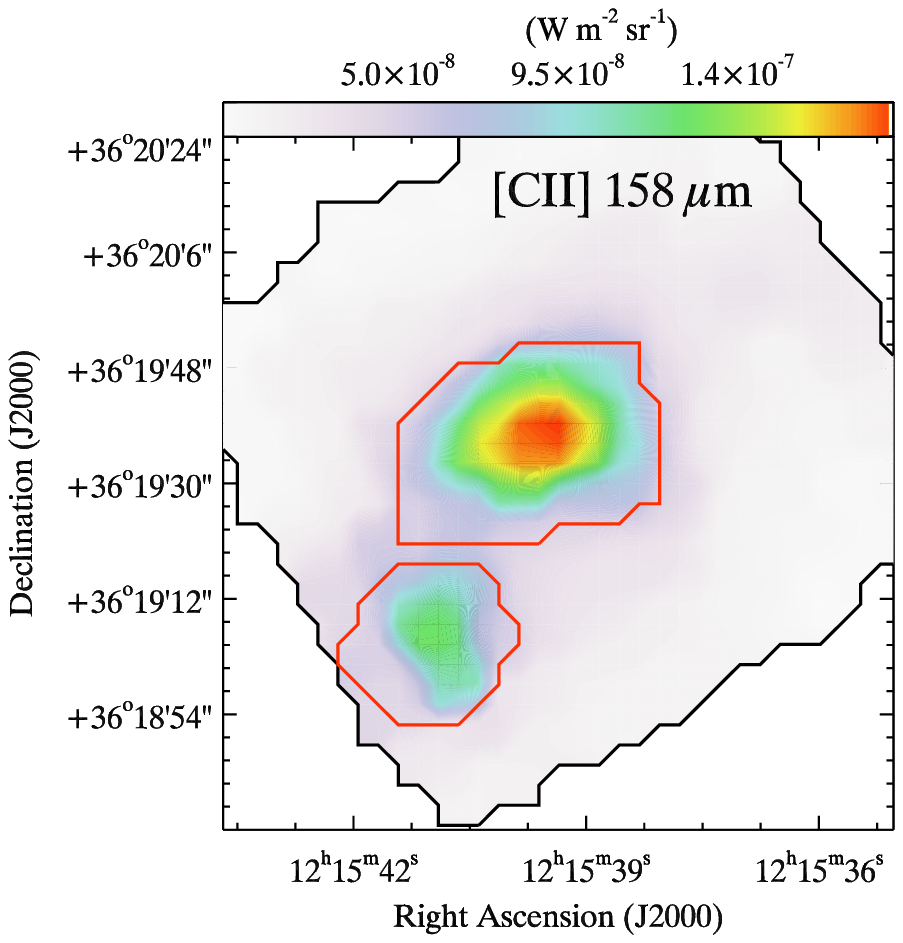}\hspace{-1mm}
\includegraphics[width=4.25cm]{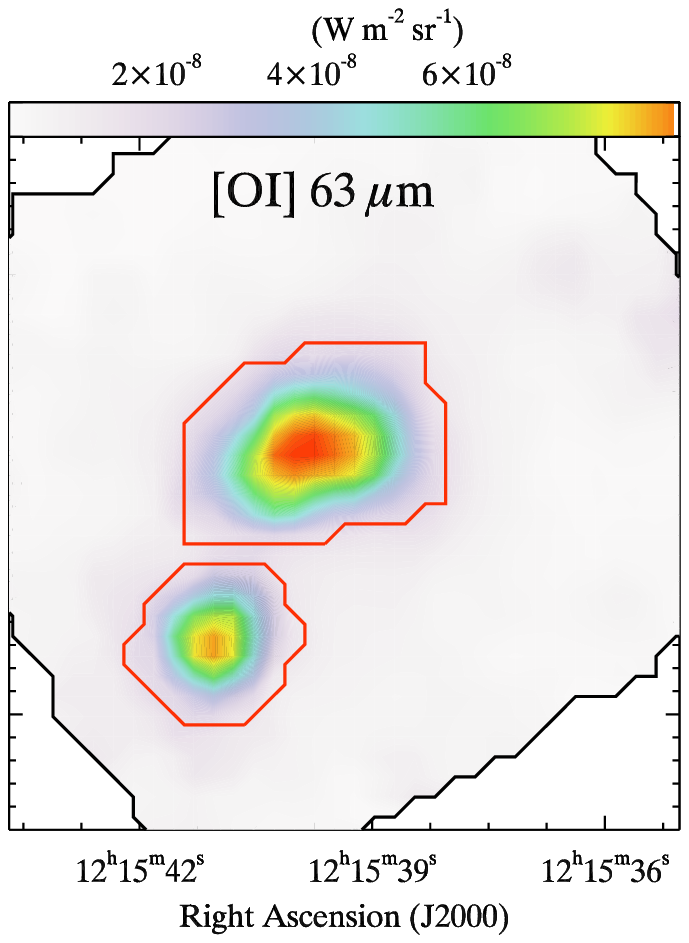}\hspace{-1mm}
\includegraphics[width=4.25cm]{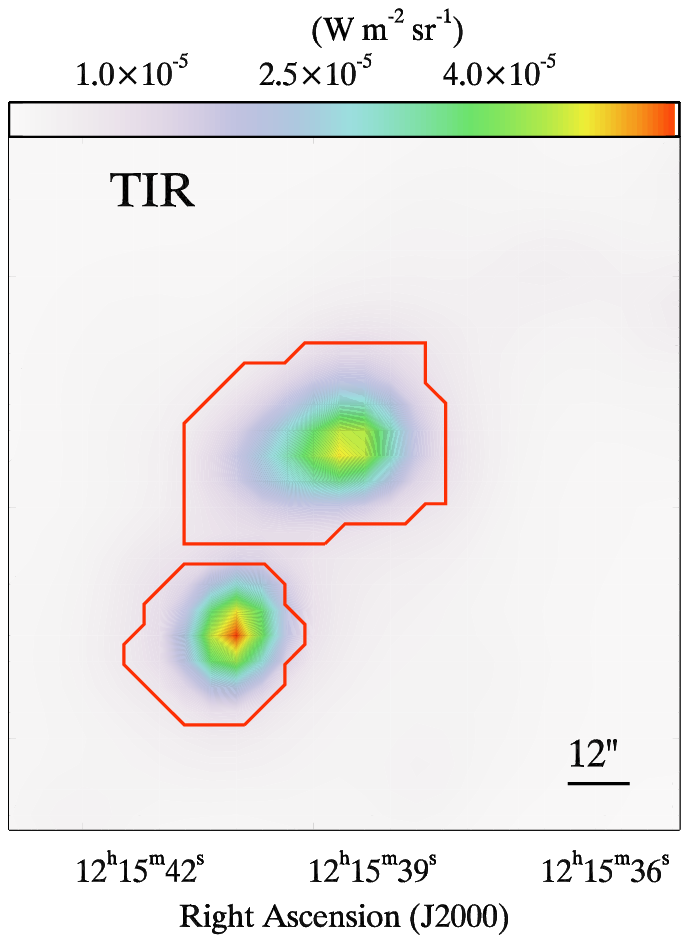}\hspace{-1mm}
\includegraphics[width=4.25cm]{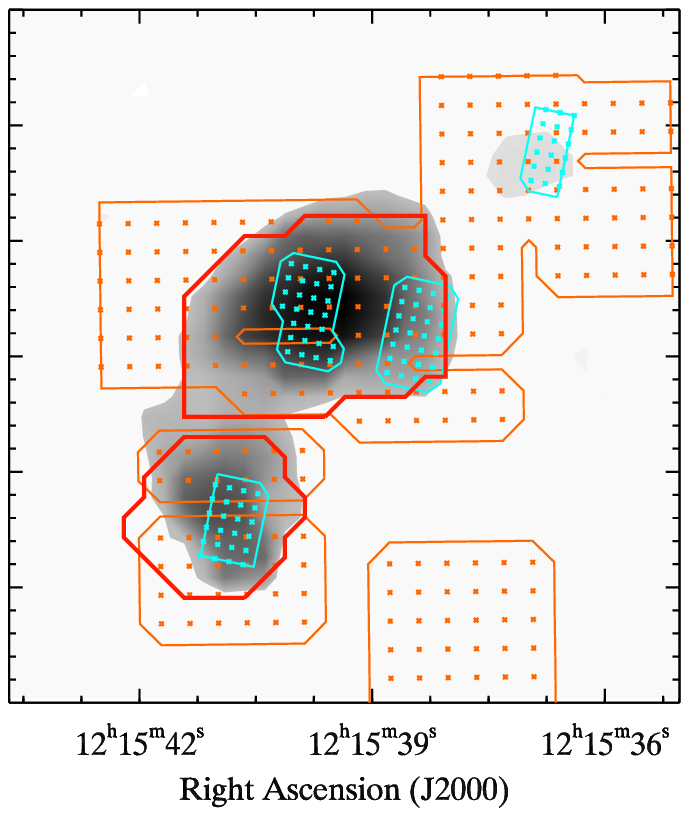}
\centering
\caption{
Maps of the \cii157\mum, \oi63\mum, 
and TIR emission in NGC\,4214. Units are W\,m$^{-2}$\,sr$^{-1}$. 
The two star-forming regions, as defined in Sect.~\ref{sect:definesf}, 
are outlined with red contours. 
The right panel shows the \spit IRS mapping strategy. 
Orange: Long-High module coverage; 
cyan: Short-High module coverage; 
gray background: \cii157\mum map. 
}
\label{mask}
\end{figure*}

We used observations of NGC\,4214 obtained by the PACS instrument \citep{poglitsch-2010} onboard the \hers Space Observatory \citep{pilbratt-2010} as part of the Dwarf Galaxy Survey \citep{madden-2013}. The list of observations can be found in Table~\ref{AOR}. The photometry data at 70\mum, 100\mum, and 160\mum, with respective beam sizes (FWHM) of 5.6\arcs, 6.7\arcs, and 11.3\arcs, were published by \cite{remy-2013}. These bands cover the peak of the SED originating from the reprocessed stellar light by the dust. 
The spectroscopy comprises observations of the \oiii88\mum and \nii122\mum lines, which trace the ionized gas, as well as the \cii157\mum, \oi63\mum, and \oi145\mum lines, which trace the PDR. The data consist of small mappings of $5\times5$ rasters separated by $\sim$16\arcs for \oiii88\mum and \oi63\mum and $3\times3$ rasters separated by $\sim$24\arcs for the other lines, ensuring a uniform coverage of $1.6^{\prime}\times1.6^{\prime}$. Originally presented in \cite{cormier-2010}, the PACS spectral data were re-processed with the reduction and analysis software HIPE user release v.11 \citep{ott-2010} and PACSman v.3.5 \citep{lebouteiller-2012}. With the improved calibration and definition of the regions, flux maps are globally consistent with those published in \cite{cormier-2010} and line ratios agree within 30\%. Flux maps of the \cii157\mum and \oi63\mum lines are shown in Fig.\,\ref{mask}. The associated error maps include data and line-fitting uncertainties, but not calibration uncertainties, which are on the order of 15\%. 
The FWHM is 9.5\arcs below 100\mum and 10\arcs, 11\arcs, 12\arcs at 122\mum, 145\mum, 160\mum, respectively. All maps were convolved to the \cii157\mum resolution of $\sim$12\arcs , which at the distance of NGC\,4214 corresponds to a physical scale of 175\,pc. 
In both the photometry and spectroscopy data sets, the convolutions were performed using kernels provided by \cite{aniano-2011}\footnote{\url{http://www.astro.princeton.edu/~ganiano/Kernels/}}.

\subsection{{\it Spitzer} data}
NGC\,4214 was observed with the three instruments onboard the \spit space telescope \citep{werner-2004}. We used the MIPS 24\mum observations obtained within the Local Volume Legacy Survey \citep{dale-2009} that were processed by \cite{bendo-2012}. The MIPS 24\mum map, which has an original FWHM of 5.9\arcs, was convolved to a resolution of $\sim$12\arcs to match that of the PACS data. \par

The IRS observations (program ID 3177, PI. Skillman) consist of small mappings of the two main star-forming regions in high-resolution mode \citep{houck-2004}. We extracted the data from the \spit Heritage Archive (see Table~\ref{AOR}) and processed them with the software CUBISM v1.8 \citep{smith-2007}. We used the default mapping procedure and bad pixel removal to produce spectral cubes with pixel sizes 2.26\arcs for the Short-High module and 4.46\arcs for the Long-High module. 
We then created surface brightness maps for all spectral lines of interest -- \siv10.5\mum, \neii12.8\mum, \neiii15.6\mum, \siii18.7\mum, \siii33.5\mum, which all trace \hii regions -- in the following way: for each pixel of the cube, we extracted the signal with a range of $\pm$0.7\mum around the line and fit a polynomial of order two for the baseline and a Gaussian for the line with the IDL routine \texttt{mpfit}. For a more stable fit, the peak of the Gaussian is required to be positive, the position of the peak is expected within one instrumental FWHM of the rest wavelength, and the width is limited to the instrumental resolution ($R=600$). Finally, we added random noise and iterated the fit $300$ times to estimate the best-fit parameters as the median of the resulting parameters and the error on those parameters as the standard deviation. 
Error maps again include data and line-fitting uncertainties, but not calibration uncertainties, which are on the order of 5\%. 
The coverage of the star-forming regions of the galaxy in the IRS maps is only partial, as shown in Fig.\,\ref{mask}. No integrated values for the flux of the whole regions could be retrieved. To obtain a representative value for the line flux in each region, we regridded the IRS maps to that of the \oiii88\mum map. Then we selected the pixels that appear in both maps and scaled the emission of these pixels to the \oiii88\mum line to infer the corresponding line fluxes for the star-forming regions as a whole. 

\begin{table}
\caption{List of \hers and \spit observations.}
\label{AOR}
 \centering
\vspace{-5pt}
  \begin{tabular}{l l}
\hline \hline
    \vspace{-10pt}\\
\multicolumn{2}{c}{\hers data}\\
Instrument & Observation Identification number (OBSID) \\
\hline
    \vspace{-10pt}\\
PACS phot. & $1342211803$, $1342211804$, $1342211805$, \\
 & $1342211806$ \\
PACS spec. & $1342187843$, $1342187844$, $1342187845$, \\
 & $1342188034$, $1342188035$, $1342188036$ \\
\hline \hline
    \vspace{-10pt}\\
\multicolumn{2}{c}{\spit data}\\
Instrument & Astronomical Observation Request (AOR) \\
\hline
    \vspace{-10pt}\\
MIPS 24\mum & $22652672$, $22652928$, $22710528$, $22710784$ \\
\multicolumn{2}{l}{IRS Short-High} \\
~On-source: & $10426368$, $10426624$, $10426880$, $10427136$ \\
~Background: & $13728256$, $13729792$, $13730304$, $13730816$, \\
& $13733120$, $13762304$, $13767424$, $13767936$, \\
& $13768448$, $13768704$, $13769728$, $13770496$, \\
& $13773568$\\
\multicolumn{2}{l}{IRS Long-High} \\
~On-source: &$10424832$, $10425088$, $10425344$, $10425600$, \\
& $10425856$, $10426112$, $10427392$\\
~Background: & $13728768$, $13729792$, $13730304$, $13730816$, \\
& $13733120$, $13763328$, $13764352$, $13765376$, \\
&$13767424$, $13767936$, $13768448$, $13768704$,\\
&  $13769728$, $13770496$, $13773568$\\
\hline \hline
\end{tabular}
\end{table}

We focus on these selected IRS lines because they are among the brightest MIR fine-structure cooling lines and can be used as reliable diagnostics of the physical conditions in \hii regions. In general, the intensity or luminosity ratio of two lines of the same element but different ionization level is indicative of the radiation field hardness. Such diagnostics are the \neiii15.6\mum/\neii12.8\mum or the \siv10.5\mum/\siii18.7\mum ratios \citep[e.g.,][]{verma-2003}, which are insensitive to the density because of their high critical densities (see Table~\ref{fluxes}). Accordingly, species of the same ionization level but different transition are indicative of the electron density as a result of the different critical densities for each transition \citep{osterbrock}. Examples are the \siii18.7\mum/\siii33.5\mum, \neiii15.6\mum/\neiii36.0\mum, or \nii122\mum/\nii205\mum ratios \citep[e.g.,][]{rubin-1994}. These diagnostics are insensitive to the temperature inside the \hii region. Unfortunately, the \neiii36.0\mum and \nii205\mum lines fall at the edge of the IRS and PACS wavelength ranges, respectively, where the spectra are too noisy to detect or derive a reliable line ratio for the two star-forming regions. Therefore we relied on the \siii line to probe the electron density.

\subsection{Total infrared luminosity map}
To construct a total infrared (TIR) luminosity map of the galaxy, we combined the MIPS 24\mum and PACS 70, 100, and 160\mum data, following \cite{galametz-2013}: 
\begin{equation}
L_{\rm TIR}=\int_{3\mu m}^{1\,100\mu m}L_{\nu}d\nu=\sum c_i L_i
.\end{equation}
We used the values of the coefficients, $c_i$, from Table~3 of their paper: $[c_{24},c_{70},c_{100},c_{160}] = [2.064,0.539,0.277,0.938]$. This method, although slightly less accurate than a direct integration of a well-sampled SED, does not require degrading the resolution of our data beyond the PACS 160\mum beam and is sufficient for our modeling purposes to estimate the energy budget in the star-forming regions. The $L_{\rm TIR}$ map is shown in Fig.\,\ref{mask}.

\subsection{Defining the star-forming regions}
\label{sect:definesf}
To define the apertures for the main star-forming regions, we set a threshold for the signal-to-noise ratio (S/N) equal to 5 in each individual PACS 70\mum, 100\mum, and 160\mum photometry and PACS spectral map. We masked all pixels below this S/N and drew the contours, which include all the remaining unmasked pixels separately for the photometry and the spectroscopy maps. Because the emission in the photometry maps is more extended, we kept the contours from the photometry and used these apertures throughout the analysis to define the two star-forming regions, as shown in Fig.\,\ref{mask}. This means that pixels in the spectroscopy maps that are below the S/N threshold but within the region contours are still counted. 
The fluxes and uncertainties for the line and TIR emission were measured taking into account all pixels in each region. They
are reported in Table~\ref{fluxes}. 
The ISM emission (gas and dust) peaks in these two regions, and most of the line fluxes are twice as high in the central region as$\text{ in}$ the southern region, except for \neiii15.6\mum (factor 1.3) and \siv10.5\mum, which have lower fluxes toward the central region. This hints at different physical conditions in the two regions, which we investigate with radiative transfer models. 

\begin{table*}
\caption{Observed MIR and FIR fluxes for the line and broadband emission.
Uncertainties on the fluxes include data and line-fitting uncertainties, but not 
calibration uncertainties, which are on the order of 5\% for the \spit lines and 15\% for the \hers lines.
Critical density and ionization potential values are taken from \citet{cormier-2012}. 
Critical densities are noted [e] for collisions with electrons and [H] for collisions with hydrogen atoms.}
\label{fluxes}
 \centering
\vspace{-5pt}
  \begin{tabular}{l c c c c c}
\hline\hline
    \vspace{-10pt}\\
 &   & \multicolumn{2}{c}{Flux $\pm$ uncertainty} & & Ionization \\ 
 Line   &  Wavelength   &       \multicolumn{2}{c}{($\times10^{-16}$ W~m$^{-2}$)} & Critical density  &  potential  \\ 
                &       (\mum)          &       Region I          &       Region II & (cm$^{-3}$) &      (eV)    \\
\hline
    \vspace{-10pt}\\
\lbrack \textsc{S\,iv}] &       10.51 & $5.68\pm0.21$           & $8.40\pm0.10$   & $5\times10^4$ [e]     & 34.79\\       
\lbrack Ne\textsc{\,ii}] &      12.81 & $8.98\pm0.22$           & $4.13\pm0.11$   & $7\times10^5$ [e]     & 21.56\\
\lbrack Ne\textsc{\,iii}] &     15.56 & $18.70\pm0.14$  & $14.25\pm0.08$        & $3\times10^5$ [e]       &40.96\\
\lbrack \textsc{S\,iii}] &      18.71 & $11.78\pm0.20$  & $6.85\pm0.08$ & $2\times10^4$ [e]       & 23.34 \\
\lbrack \textsc{S\,iii}] &      33.48 & $18.71\pm0.27$  & $8.20\pm0.12$ & $7\times10^3$ [e]       & 23.34\\
\lbrack \textsc{O\,i}] &        63.18 & $10.11\pm0.35$  & $4.06\pm0.21$ & $5\times10^5$ [H]       & -\\
\lbrack \textsc{O\,iii}] &      88.36 & $31.86\pm0.62$  & $13.50\pm0.40$        & $5\times10^2$ [e]       & 35.12\\
\lbrack \textsc{N\,ii}] &       121.90 &        $0.44\pm0.20$           & $0.14\pm0.08$   & $3\times10^2$ [e]     & 14.53\\ 
\lbrack \textsc{O\,i}] &        145.52 &        $0.65\pm0.09$           & $0.32\pm0.07$   & $1\times10^5$ [H]     & -\\
\lbrack \textsc{C\,ii}] &       157.74 &        $26.34\pm0.33$  & $10.05\pm0.21$  & 50 [e], $3\times10^3$ [H]     & 11.26\\
\hline
\end{tabular}
\begin{tabular}{l c c c}
\hline
    \vspace{-10pt}\\
   &  & \multicolumn{2}{c}{Flux density $\pm$ uncertainty} \\
 Broadband      &       Wavelength      & \multicolumn{2}{c}{(Jy)}\\
                &               (\mum)  &       Region I          &       Region II              \\
\hline
    \vspace{-10pt}\\
MIPS  & 24 & $0.67\pm0.01$ & $0.48\pm0.01$ \\
PACS & 70 & $7.36\pm0.21$ &     $3.72\pm0.12$ \\
PACS & 100 & $7.91\pm0.19$ & $4.10\pm0.11$ \\
PACS & 160 & $6.07\pm0.10$ & $3.23\pm0.06$ \\
\hline
    \vspace{-10pt}\\
{$L_{\rm TIR}$ (erg\,s$^{-1}$)} & 3 -- 1100 & $5.25\times10^{41}$ &       $3.02\times10^{41}$     \\
\hline \hline
\end{tabular}
\end{table*}

\section{Description of the model}
\label{method}
\subsection{Model geometry and strategy}
Our objective is to characterize the physical conditions of the ISM phases from which the IR emission arises in NGC\,4214. To that end, we used the spectral synthesis code \textsc{Cloudy} v.13, last described by \cite{ferland-2013}. We performed a multiphase detailed modeling of the ISM for which we combined line and continuum emission, following the method described in \cite{cormier-2012}. We considered the two main star-forming regions of NGC\,4214: the most evolved central region (NW-I) and the southern region (SE-II). Here we present the main aspects of the model and how it is applied to each region. The components/ISM phases of the model are 
\begin{enumerate}
         \item a central source of radiation,
         \item an ionized medium component (\hii region) that surrounds the central source, 
         \item a neutral medium (PDR) surrounding the \hii region.
\end{enumerate}
\noindent 
This method assumes a single radiation source responsible for the observed SED of the studied region. In other words, we took all of the different sources (star clusters) and the surrounding clouds from which they have formed and represented them with one central source and one surrounding cloud. We thus targeted the integrated properties of each region. In practice, we mixed components of the ISM that have different composition and properties
and blend them in a single system. 
The applied geometry is spherical. The source is in the center and the illuminated face of the cloud lies at a certain distance that we call inner radius. In our case, the effective geometry is one-dimensional plane-parallel because the cloud forms a thin shell and its distance from the radiation source is large. \par

The radiation source, representative of the stars that populate the clusters of the star-forming region, illuminates a cloud of dust and gas. It controls the ionization parameter, $U$, which characterizes the field and is defined as the ratio of the incident ionizing photon density to the hydrogen density. Hard UV photons from the source ionize hydrogen and form the \hii region. As this radiation is transmitted through the cloud, it is attenuated
and thus becomes softer, which decreases its influence on ionization. However, it still controls the processes further in the cloud (in the PDR).\par

The adopted strategy is to treat the \hii region first and then use the \hii region parameters as input for the PDR modeling. This allows for a self-consistent approach \citep{abel-2005}, which is usually not directly available in standard PDR codes (see \citealt{roellig-2007} for a comparison of PDR codes), and is important to accurately derive the radiation field that impinges
on the PDR. We first ran the simulation until the end of the \hii region, choosing to stop the simulation when the (electron) temperature reached 500\,K to ensure that the model had transitioned to the atomic phase. We iterated to optimize our parameters so
that they matched the observed emission of known \hii region-diagnostic lines ([\textsc{S\,iv}]\,10.5\mum, [Ne\textsc{\,ii}]\,12.8\mum, [Ne\textsc{\,iii}]\,15.6\mum, [\textsc{S\,iii}]\,18.7\mum, [\textsc{S\,iii}]\,33.5\mum, \oiii88\mum, and \nii122\mum). Then we fed the result of this model to the PDR and compared the predictions to the remaining three PDR lines observed: \oi63\mum, \oi145\mum, and \cii157\mum, choosing a visual extinction of 10\,mag as the stopping criterion. At this point, the gas temperature had fallen to roughly 10\,K.

\subsection{Model parameters}
We constrained the properties of the star-forming regions by varying some of the parameters that control the physics of the models while keeping others fixed. The main parameters that we consider are
\begin{enumerate}
  \item a radiation field source: shape, age, luminosity (varied),
  \item the hydrogen density of the ISM, $n_{\rm H}$ (varied),
  \item the ISM gas elemental abundances (fixed),
  \item the inner radius, $r_{\rm in}$ (varied),
  \item the magnetic field, B (fixed),
  \item the turbulent velocity, $v_{\rm turb}$ (fixed).
\end{enumerate}
\noindent Parameters that are fixed were set to values from the literature. The other parameters were varied inside a range whose width reflects the dispersion of published measurements or of the data. The main parameter of interest for this study is the radiation field, which was varied within a range guided by studies of the star formation history.

\subsubsection{Hydrogen density ($n_{\rm H}$)}
\label{density}
We performed our simulations assuming pressure equilibrium. As the model proceeds through consecutive zones of the cloud, it keeps the pressure constant. Thus, the density of the medium varies to satisfy this equilibrium. The initial density that we specified in the models is the density at the illuminated face of the cloud, where the \hii region starts. The initial values and the range we probed are motivated by the observed \siii18.7\mum/\siii33.5\mum ratio in the \hii region, which is
known to be sensitive to the electron density in the range 10$^2$-10$^4$\,cm$^{-3}$ \citep{rubin-1994}. We therefore let the initial density vary in the range 100-300\,cm$^{-3}$ for the central region and 300-600\,cm$^{-3}$ for the southern region with a common step of 25\,cm$^{-3}$. By iterating this procedure, we constrained the best values for the density at the beginning of the \hii region.

\subsubsection{Inner radius ($r_{\rm in}$)}
In our spherical geometry, the source is at the center and is surrounded by a cloud. The illuminated face of the cloud lies at a certain distance $r_{\rm in}$. This is not a strictly physically constrained parameter because the setup we used does not realistically model each cluster, but instead tries to mimic a whole region and reproduce its emission. The variation of this radius changes the photons flux and thus is expected to affect our results. We let $r_{\rm in}$ vary from 1 to 100\,pc for both regions.

\subsubsection{Elemental abundances}
Elemental abundances in the models were set to the observed values for oxygen, sulfur, nitrogen, and neon, taken from \cite{kobulnicky-1996}. Some measurements partially cover our defined regions, and we adopted them as representative. Exclusively for carbon, we scaled its abundance according to the study on the dependence of $\log(C/O)$ on metallicity by \cite{izotov-1999}. For other elements, we used the default ISM composition of \textsc{Cloudy} and scaled the abundances to our metallicity (1/3). The values used are indicated in Table~\ref{elements}. 

\begin{table}
  \caption{Elemental abundances in NGC\,4214. 
  Values (in logarithmic scale) for NGC\,4214 are taken from \citet{kobulnicky-1996} and solar values from \citet{asplund-2009}.}
\label{elements}
 \centering
\vspace{-5pt}
   \begin{tabular}{c c c c}
   \hline 
    \vspace{-10pt}\\
        Abundance               & Region I      & Region II  & Solar value \\
        \hline
    \vspace{-10pt}\\
        \lbrack O/H]            & $-3.795\pm0.05$       & $-3.64\pm0.04$        & $-3.31\pm0.05$ \\
        \lbrack S/H]            & $-5.380\pm0.06$       & $-5.21\pm0.06$        & $-4.88\pm0.03$ \\
        \lbrack N/H]            & $-5.094\pm1.00$       & $-5.02\pm0.10$        & $-4.17\pm0.05$ \\
        \lbrack Ne/H]           & $-4.535\pm0.11$       & $-4.51\pm0.08$        & $-4.07\pm0.10$ \\
        \lbrack C/H]            & $-4.295\pm0.30$       & $-4.14\pm0.30$        & $-3.57\pm0.05$ \\
        \hline 
   \end{tabular}
 \end{table}

\subsubsection{Radiation field - shape and energy}
We used the code \textsc{Starburst99}  \citep{leitherer-2010} to produce a stellar population spectrum that serves as input for our models. We chose a Kroupa initial mass function between 0.08 and 120\,M$_{\odot}$ \citep{kroupa-2001}, as done in \cite{andrews-2013}, and Padova asymptotic giant branch tracks with Z=0.004. 
As discussed above, we did not model each cluster individually, but we used integrated emission from the entire star-forming regions instead. We tried to be as close to the shape and intensity of the radiation field of the \hii regions as possible. Motivated by the star formation history of the galaxy presented in \cite{mcquinn-2010} and \cite{williams-2011}, we tested the following cases:
\begin{itemize}
\item For the central region, we considered two limiting scenarios: 
(i)~a single-burst star formation event and (ii)~a continuous star formation model, with SFR=0.07\,${\rm M_{\odot}\,yr^{-1}}$. The ages of the clusters were varied within a range of (i)~1-20\,Myr (in steps of 0.5\,Myr) and (ii)~200-1000\,Myr (in steps of 200\,Myr).
\item For the southern region, we considered a single-burst event with an age that varied from 1 to 20\,Myr (in steps
of 0.5\,Myr).
\end{itemize}
\noindent
A fixed mass of 10$^5$\,M$_{\odot}$ (typical cluster mass in \citealt{sollima-2013}) was considered for the single bursts, where the stars are created at once (delta burst). Our value for the SFR (0.07\,${\rm M_{\odot}\,yr^{-1}}$) is representative of the `average' rate at which this galaxy formed stars within the past 1\,Gyr of its history. The ages of the clusters in both regions were guided by values from \cite{ubeda-2007}, \cite{sollima-2013}, and \cite{sollima-2014}. \cite{ubeda-2007} found 2-7\,Myr for region~I, along with extended clusters in the same region with ages of 150 to 190\,Myr. In region~II, they found ages spreading around 2\,Myr. \cite{sollima-2013,sollima-2014} reported a larger spread in ages. In region~I, they obtained ages around a median of 14\,Myr, and for the more extended clusters the ages lie between 10 and 300\,Myr. For region~II, they found a median age of $\sim$20\,Myr. \par

For the luminosity emitted from each region, we chose to use the TIR luminosity as a first approximation of the luminosity of the starburst. This choice implies the assumption that all radiation from the clusters in the region is reprocessed by the dust and thus emitted at longer wavelengths. In doing so, we kept in mind that there can be processes that we did not model (UV escape fraction or a diffuse ionized medium, for example) and that can contribute to this radiated energy (see Sect.~\ref{discussion}).

\subsubsection{Magnetic field strength (B) and turbulent velocity ($v_{\rm turb}$)}
Magnetic fields and turbulence play an important role in structuring the ISM \citep[e.g.,][]{mckee-2007}. When \textsc{Cloudy} solves the pressure equilibrium for each zone of the modeled cloud, a magnetic pressure term equal to $\displaystyle P_{\rm B}=\frac{B^2}{8\pi}$ is included in the equation of state along with a turbulent pressure term equal to $\displaystyle P_{\rm turb}=2.8 \cdot 10^6 \cdot 3 \cdot (\frac{n_{\rm H}}{10^5\,{\rm cm^{-3}}})(\frac{v_{\rm turb}}{\rm {1\,km\,s^{-1}}})^2$~[cm$^{-3}$\,K], for isotropic turbulent motions, where $n_{\rm H}$ is the total hydrogen density and $v_{\rm turb}$ is the turbulent velocity (see the {\sc Hazy} documentation of \textsc{Cloudy} for more information). \par

The magnetic field of NGC\,4214 was measured by \cite{kepley-2011} using multiwavelength radio emission. The reported field strength in the center of the galaxy is 30\,$\mu$G, and the pressure term due to this field has the same order of magnitude as the hot gas and the gravitational contributions. Since it is not well known how the observed magnetic field might affect our observed line intensities, we excluded it from our default models and tested one case with a magnetic field strength of 30\,$\mu$G. \par 

Another potential energy source to consider can arise from the dissipation of turbulence. Turbulent energy is converted into thermal energy as it cascades from large scales to small scales through dissipation. However, we did not resolve size scales for which we can measure this. 
\textsc{Cloudy} does not attempt to model the dissipation mechanism, but assumes a simple thermal energy source based on line width. The turbulent velocity was set to a value of 1.5\,km\,s$^{-1}$ by default in our models, and we tested two other cases: one case with an intermediate turbulent velocity ($v_{\rm turb}$=3\,km\,s$^{-1}$ or FWHM=5\,km\,s$^{-1}$) that corresponds to the approximate line width observed in the CO(1-0) data by \cite{walter-2001}, and one case with a high turbulent velocity ($v_{\rm turb}$=50\,km\,s$^{-1}$) as found in the diffuse ionized gas by \cite{WilcotsThurow-2001} and used also in \cite{kepley-2011}.

Nevertheless, we explore the effects of excluding or including magnetic fields and turbulence in Sect.~\ref{magn}.

\subsection{Determination of best-fitting models}
\label{minchi}
We aim to converge on a unique parameter set that best describes the conditions of the regions. We first ran models for which
we varied the parameters in a coarse grid to narrow down the parameter space, using ranges of values found in the literature to start with. We then used the \textit{optimization} option of \textsc{Cloudy}, which automatically varies the specified parameters in a finer grid to find the optimal solution.

We computed the average $\chi ^2$, denoted $\bar\chi ^2$, for each model by comparing the observed fluxes of \siv10.5\mum, \neiii15.6\mum, \siii18.7\mum, \siii33.5\mum, and \oiii88\mum to the fluxes predicted from the radiative transfer calculation. These lines are the most luminous and most strongly correlated with the \hii region (as opposed to \nii and \neii which, from experience, can arise from other phases). We refer to these as the optimized lines in Table~\ref{chi2ionic}. The goodness of the line emission fit is given by low $\bar\chi ^2$ values.
The \textit{optimization} method of \textsc{Cloudy} searches the minimum of $\bar\chi ^2$ that is defined as
\begin{align}
   \bar\chi ^2=\frac{1}{n}\sum{\chi_i ^2}=\frac{1}{n}\sum{\frac{( M_i - O_i)^2}{(min\{O_i;M_i\} \times \sigma_i)^2}},
\end{align}
\noindent where $n$ is the number of lines optimized and $\chi_i^2$ are the $\chi^2$ values of the individual optimized lines.
M$_i$ and O$_i$ are the modeled and observed fluxes, and $\sigma_i$ is the fractional error on the observed flux (uncertainty/flux) with calibration uncertainties added in quadrature to the measured uncertainties described in Sect.~\ref{data}. We have five observables (the ionic lines listed above) and varied four parameters (cluster age, source luminosity, hydrogen density, and inner radius).

\section{Results}
\label{results}
In this section we present the model results for the two regions according to the star formation histories considered. Parameters of the best-fitting models and their corresponding $\chi^2$ values are reported in Tables~\ref{model} and \ref{chi2ionic}, respectively.

\begin{table*}[!t]
\caption{Input parameters for the best-fitting models. 
Values are given at the illuminated face of the cloud. 
Values in parenthesis for the magnetic field strength 
and turbulent velocities are tested in Sect.~\ref{magn}.}
\label{model}
  \centering
\vspace{-5pt}
  \begin{tabular}{l c c c}
  \hline 
    \vspace{-10pt}\\
  Parameter                             & \multicolumn{2}{c}{Central region} & Southern region\\
                                                & Single burst & Continuous & Single burst \\
  \hline
    \vspace{-10pt}\\
  Density $n_{\rm H}$ [cm$^{-3}$]               & 155           & 180             & 440   \\
  Inner radius $r_{\rm in}$ [pc]                        & 85.4                 & 62.1          & 22.3  \\
  Stellar age t [Myr]                                   & 4.1                 & 440           & 3.9            \\
  Total luminosity L [erg\,s$^{-1}$]            & $1.15\times10^{42}$ & $1.91\times10^{42}$   & $5.25\times10^{41}$   \\
  Magnetic field B [$\mu$G]                     & -                     & - (30)          & -             \\
  Turbulent velocity $v_{\rm turb}$ [km\,s$^{-1}$]      & 1.5         & 1.5 (3, 50)   & 1.5           \\
  \hline 
  \end{tabular}
\end{table*}
\begin{figure*}
\centering
\includegraphics[clip,trim=0 20mm 0 0,width=17cm,height=5cm]{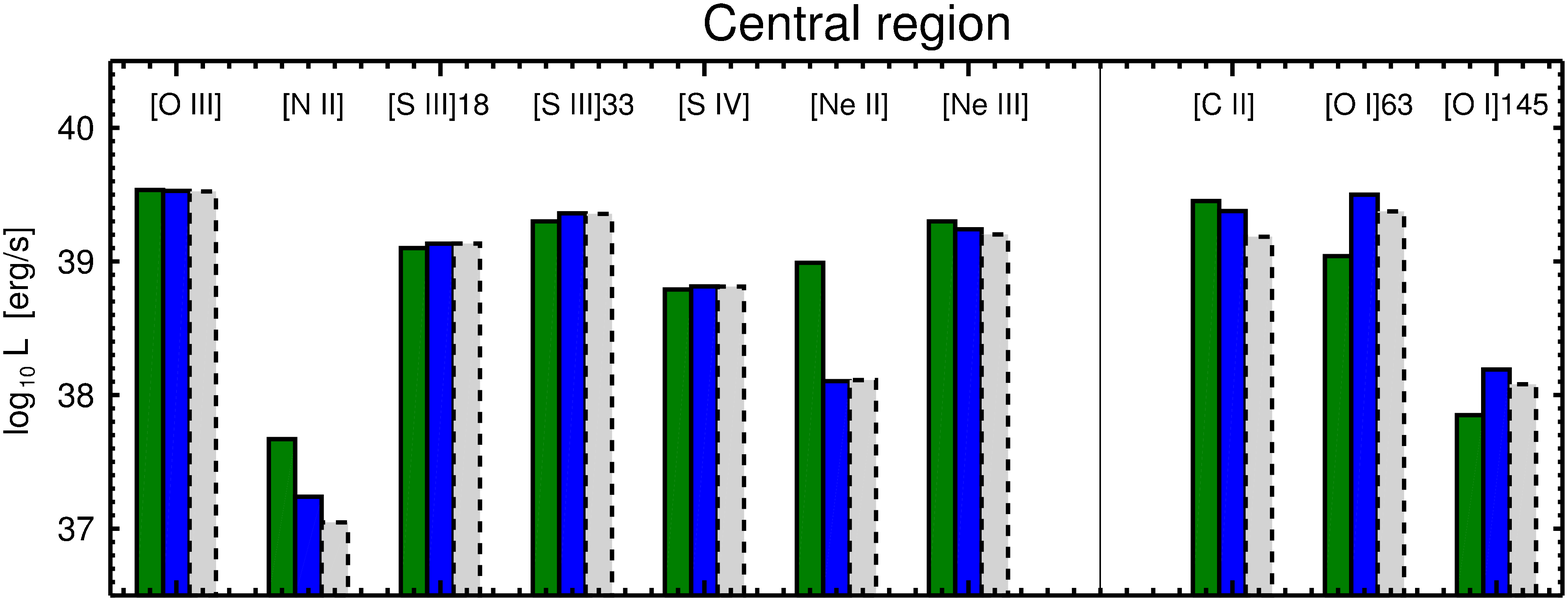}
\includegraphics[clip,trim=0 20mm 0 0,width=17cm,height=5cm]{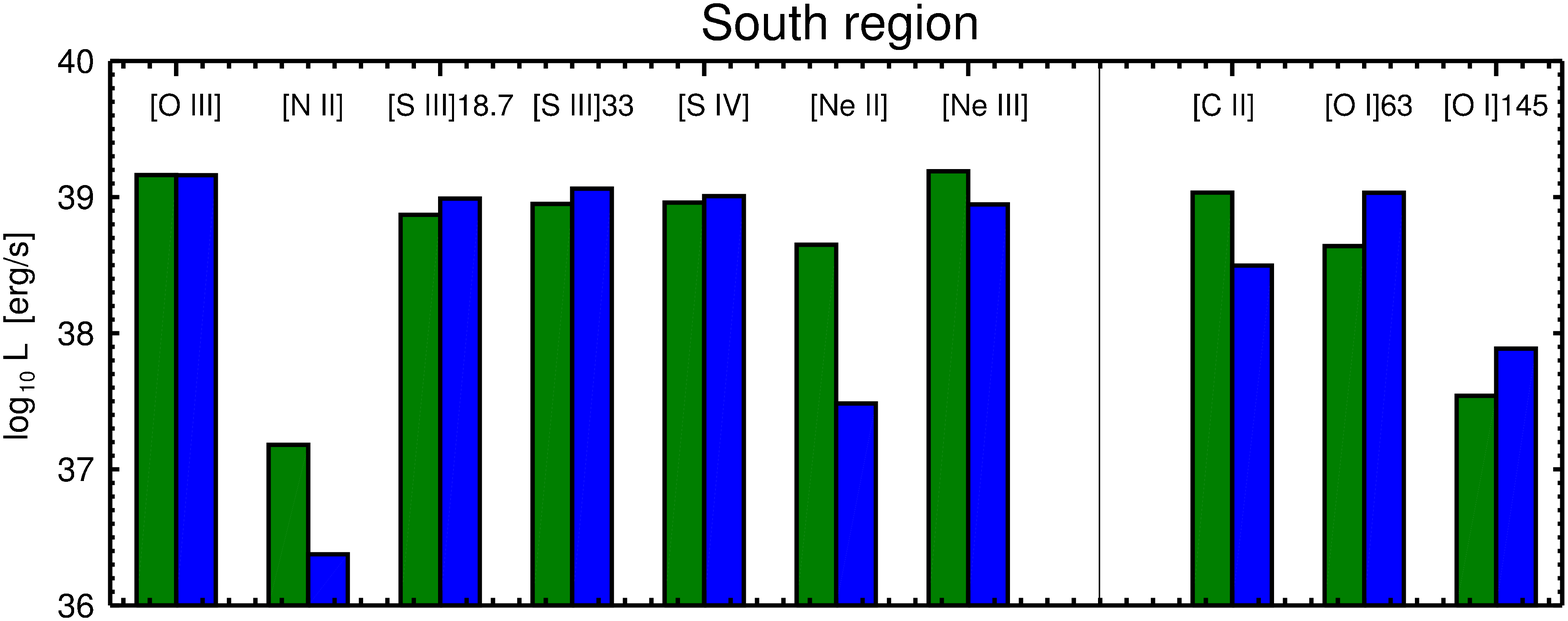}
\caption{Results for the central region (top panel) 
and for the southern region (bottom panel): line emission 
for the \hii region (left side) and the PDR (right side), assuming 
pressure equilibrium.
Green bars represent the observations, blue bars our single-burst model predictions, 
and gray bars with a dashed outline the continuous star formation model predictions.}
\label{h2bars}
\end{figure*}

\subsection{Line emission}
\label{modelresults}
\subsubsection{Central region (I): The single-burst model}
\label{SB}
For the single-burst star formation event, the best-fitting model of the central region has the following parameters: burst age of 4.1\,Myr with luminosity $1.1\times10^{42}$\,erg\,s$^{-1}$, density of 155\,cm$^{-3}$, and $r_{\rm in}\simeq85$\,pc. The corresponding ionization parameter at the illuminating face of the cloud is $\log(U)=-2.7$. For the \hii region, all optimized lines are matched within $\pm$30\%, and \neii12.8\mum and \nii122\mum are underpredicted by a factor of $\sim$8 and 3, respectively (see Fig.\,\ref{h2bars}, blue bars in the top panel). 
In the PDR, the \oi145\mum and 63\mum lines are overpredicted by a factor of $\sim$2.5, and the \cii157\mum line is matched within 20\%. We note that \cii emission can arise from both the neutral and the ionized phases of the ISM, with a potentially non-negligible contribution from the warm ionized medium in the Milky Way \citep{heiles-1994}. The \hii region of our best-fitting model contributes negligibly to the predicted \cii157\mum and \oi emission ($<$3\%). The ultraviolet radiation field strength $G_0$ inside the PDR is $455$ in units of the equivalent \cite{habing-1968} flux ($1\,G_0 = 1.6\times10^{-3}$\,erg\,cm$^{-2}$\,s$^{-1}$). The input luminosity required by the model, which all comes out as $L_{\rm TIR}$ in the PDR, is $\text{about twice as}$  high
as the observed $L_{\rm TIR}$ in that region. 

These results represent the simplifying case that the recent starburst dominates the star formation in this central region, so the line emission can be explained with a single burst. The age of the burst in the model nicely agrees with the range of ages from \cite{ubeda-2007} and is at the younger end of ages from \cite{sollima-2013,sollima-2014}.

\subsubsection{Central region (I): The continuous star formation model}
In the continuous star formation scenario, the best-fitting model of the central region has the following parameters: stellar age of 440\,Myr with luminosity $1.9\times10^{42}$\,erg\,s$^{-1}$, density of 180\,cm$^{-3}$, and $r_{\rm in}\simeq62$\,pc. The ionization parameter is $\log(U)=-2.5$. Line predictions for the \hii region and the PDR from this model solution are shown in Fig.\,\ref{h2bars} (gray bars). For all optimized lines (\oiii88\mum, \siii18.7 and 33.5\mum, \siv10.5\mum, and \neiii15.6\mum), the model matches the observations within $\pm$20\%. The two other ionic lines, \neii12.8\mum and \nii122\mum, are underpredicted by a factor of $\sim$7 and 4. 
In the PDR, \cii and the \oi lines are matched within a factor of $\sim$2. The contribution of the \hii region to the predicted PDR emission is only 1\%. 
We find $G_0\simeq1.2\times10^3$, which is higher than in the single-burst case. The luminosity of the model exceeds the observed $L_{\rm TIR}$ by a factor of 3.5. \par

The results represent the simplifying case of continuous star formation dominating this region, with the starbursts being embedded in it. The age of the model agrees well with the star formation event in the window 400-500\,Myr ago reported by \cite{mcquinn-2010}.

\subsubsection{Southern region (II)}
\label{southres}
The best-fitting model for the southern region is characterized by a burst age of 3.9\,Myr with luminosity $5.3\times10^{41}$\,erg\,s$^{-1}$, density of 440\,cm$^{-3}$, and $r_{\rm in}\simeq22$\,pc. The ionization parameter is $\log(U)=-2.3$. Line predictions for the \hii region and the PDR are shown in the bottom panel of Fig.\,\ref{h2bars}. This burst found for the southern region is slightly younger than the burst in the central region, in agreement with \cite{ubeda-2007}. In the \hii region, the \oiii88\mum, \siii18.7 and 33.5\mum, and \siv10.5\mum lines are reproduced within 30\%, while \neiii15.6\mum, \neii12.8\mum, and  \nii122\mum are underpredicted by a factor of 1.7, 10, and 6, respectively. 
Feeding this model to the PDR, the \cii157\mum line is underpredicted by a factor of 3.4, while the \oi lines are both overpredicted by a factor of $\sim$2. The contribution of the \hii region to the PDR line emission is only 1\%. $G_0$ is found to be about $3.2\times10^3$, which is higher than in the central region. The luminosity of the model is 1.7 times higher than the observed $L_{\rm TIR}$ in this region.

\subsubsection{Comparison to empirical line ratios}
Physical conditions in the \hii region are mainly determined by tracers of the radiation field strength (\neiii/\neii, \siv/\siii18.7\mum) and of density (\siii18.7\mum/\siii33.5\mum). We compare these well-known diagnostic ratios in the two star-forming regions in Table~\ref{ratios}. In the southern region, ratios of \neiii/\neii and \siv/\siii18.7\mum are observed to be about twice as high and \siii18.7\mum/\siii33.5\mum marginally higher than in the central region, indicating that the radiation field is harder and the medium denser. This is indeed what we recover with our best-fitting models (Table~\ref{model}), as they match the sulfur ratios well.

\begin{table}
\caption{$\chi^2$ values for the two star-forming regions.}
\label{chi2ionic}
 \centering
\vspace{-5pt}
   \begin{tabular}{l c c c}
   \hline \hline
    \vspace{-10pt}\\
        \hii region~~~  &\multicolumn{2}{c}{Central region}    & \hspace{-2mm}Southern region \\             
                                & Burst         & \hspace{-1mm}Continuous & Burst \\
        \hline
    \vspace{-10pt}\\
        \multicolumn{2}{l}{Individual $\chi^2$ values:}         &                 &               \\
        \lbrack \textsc{O\,iii}] \,88\mum               & 0.01  & 0.03    & 0.01  \\
        \lbrack \textsc{N\,ii}]\,122\mum                & 12.58 & 44.63   & 82.56 \\
        \lbrack \textsc{S\,iii}]\,18.7\mum              & 2.24  & 2.38    & 37.69 \\
        \lbrack \textsc{S\,iii}]\,33.5\mum              & 8.10  & 6.99    & 31.89 \\
        \lbrack \textsc{S\,iv}]\,10.5\mum       & 0.76  & 0.70  & 4.94    \\
        \lbrack Ne\textsc{\,ii}]\,12.8\mum      & 14442.2       & 13842.6 & 58101.7       \\
        \lbrack Ne\textsc{\,iii}]\,15.6\mum     & 8.14  & 24.65         & 224.51        \\
        \hline
    \vspace{-10pt}\\
        $\bar\chi^2$ (all ionic lines)          & 2067.72       & 1988.85 & 8354.76       \\
        $\bar\chi^2$ (optimized lines) \hspace{-4mm}          & 3.85  & 6.95  & 59.81   \\
        \hline
        \hline
    \vspace{-10pt}\\
        PDR~~~  & \multicolumn{2}{c}{Central region}    & \hspace{-2mm}Southern region \\               
                                & Burst         & \hspace{-1mm}Continuous & Burst \\
        \hline
    \vspace{-10pt}\\
        \multicolumn{2}{l}{Individual $\chi^2$ values:}         &                 &               \\
        \lbrack \textsc{C\,ii}]\,157\mum        & 1.57  & 31.83 & 258.63  \\
        \lbrack \textsc{O\,i}]\,63\mum  & 149.78        & 57.05 & 85.37   \\
        \lbrack \textsc{O\,i}]\,145\mum & 34.14 & 11.83 & 21.27 \\
        \hline
    \vspace{-10pt}\\
        $\bar\chi^2$ (all PDR lines)    & 61.83 & 33.57 & 121.76        \\
        \hline \hline
   \end{tabular}
 \end{table}
\begin{table*}
\caption{Observed and predicted MIR line ratios for the two star-forming regions.
The \neiii/\neii and \siv/\siii line ratios are indicative of the radiation field strength, 
and the ratio of the two \siii lines is a density diagnostic.}
\label{ratios}
\centering
\vspace{-5pt}
   \begin{tabular}{l c c c c c c c}
   \hline
    \vspace{-10pt}\\
Ratio           & & \multicolumn{3}{c}{Central region}          & & \multicolumn{2}{c}{Southern region} \\
   & &   Observed &    Burst & Continuous                        & & Observed      & Burst \\
   \hline
    \vspace{-10pt}\\
\lbrack Ne\textsc{\,iii}]/\neii& & 2.082        & 13.72 & 12.34          & & 3.448               &       28.955          \\  
\lbrack \textsc{S\,iv}]/[\textsc{S\,iii}]18.7 & &       0.481 & 0.479 & 0.477                 & & 1.227  &  1.044\\
\lbrack \textsc{S\,iii}]18.7/[\textsc{S\,iii}]33 & &  0.630  & 0.593         & 0.601         & & 0.835 & 0.846\\
\hline
   \end{tabular}
  \end{table*}

\subsection{Effects of magnetic fields and turbulence on the PDR temperature and density}
\label{magn}
\begin{figure}
\centering
\includegraphics[clip,width=8.2cm]{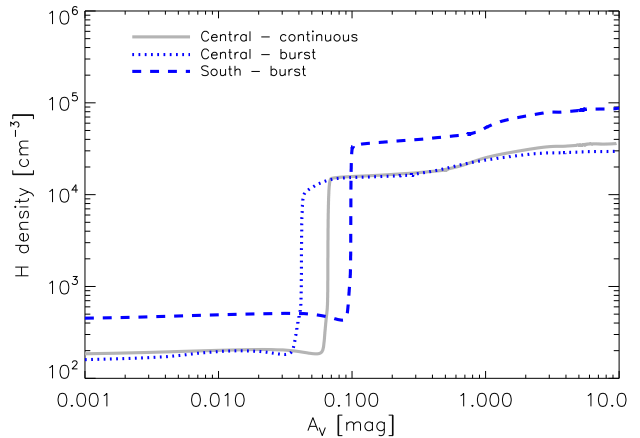}
\vspace{-5pt}
\caption{Density profiles in the modeled clouds for the central and southern region, which include a turbulence pressure term ($v_{\rm turb}$=1.5\,km\,s$^{-1}$). Note that the x-axis is logarithmic, so the \hii region (with a constant low density) occupies a thin layer of the cloud, stopping at low visual extinction ($\sim$0.1\,mag).}
\label{dens}
\end{figure}

After determining the best parameters for the \hii regions, we explored the effect of cloud density on the PDR emission in more
detail. 
Density is of critical importance in the emission output of the simulation because of the different critical densities of the observed lines. Density values quoted so far are representative of the \hii region and evolve inside the modeled cloud. Figure~\ref{dens} shows hydrogen density profiles in the clouds for each case presented in Sect.~\ref{modelresults}. The density starts at the initial value we set for each model in the \hii region, remaining practically at the same level throughout the \hii region. At the interface between the \hii region and the PDR, there is a jump in density required to keep the model in pressure equilibrium. When pressure is only determined by the gas pressure ($P_{\rm gas}=n_{\rm H}kT$; i.e., magnetic fields and turbulence are omitted), the temperature difference at the phase transition is balanced by a rise in density of 2-3 orders of magnitude. Within this frame, the effects of magnetic fields or turbulence, implemented as pressure terms in \textsc{Cloudy}, can be understood as follows: when total pressure equilibrium is assumed, they give more support at the phase transition, thus preventing a large difference in density between the \hii region and the PDR and moderating the density increase at large optical depths. 
For the model presented in Sect.~\ref{modelresults}, where a low turbulence value is included but no magnetic fields, representative PDR densities are $2\times10^4$\,cm$^{-3}$ in the central region for both the single-burst and continuous star formation models, and $7\times10^4$\,cm$^{-3}$ in the southern region (see Fig.\,\ref{dens}). 

\begin{figure*}
\centering
\includegraphics[clip,width=\textwidth,height=6cm]{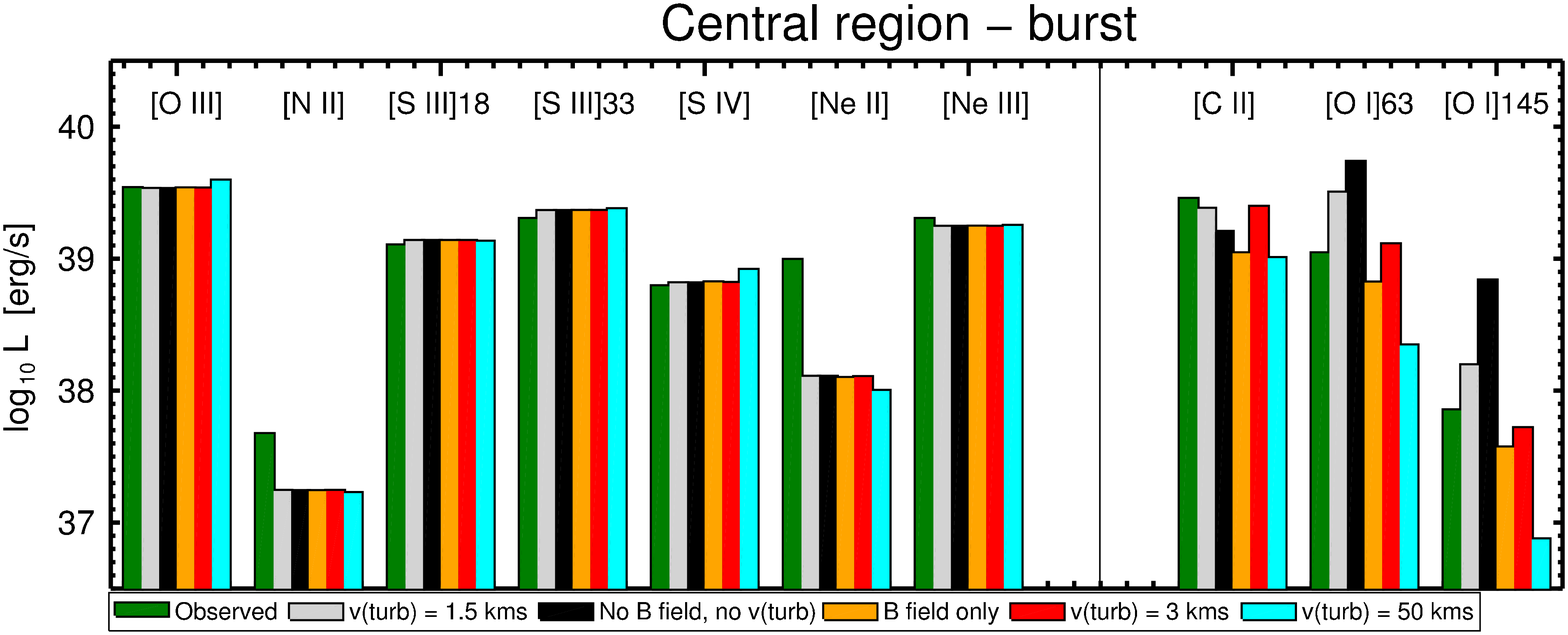}
\includegraphics[clip,width=8cm]{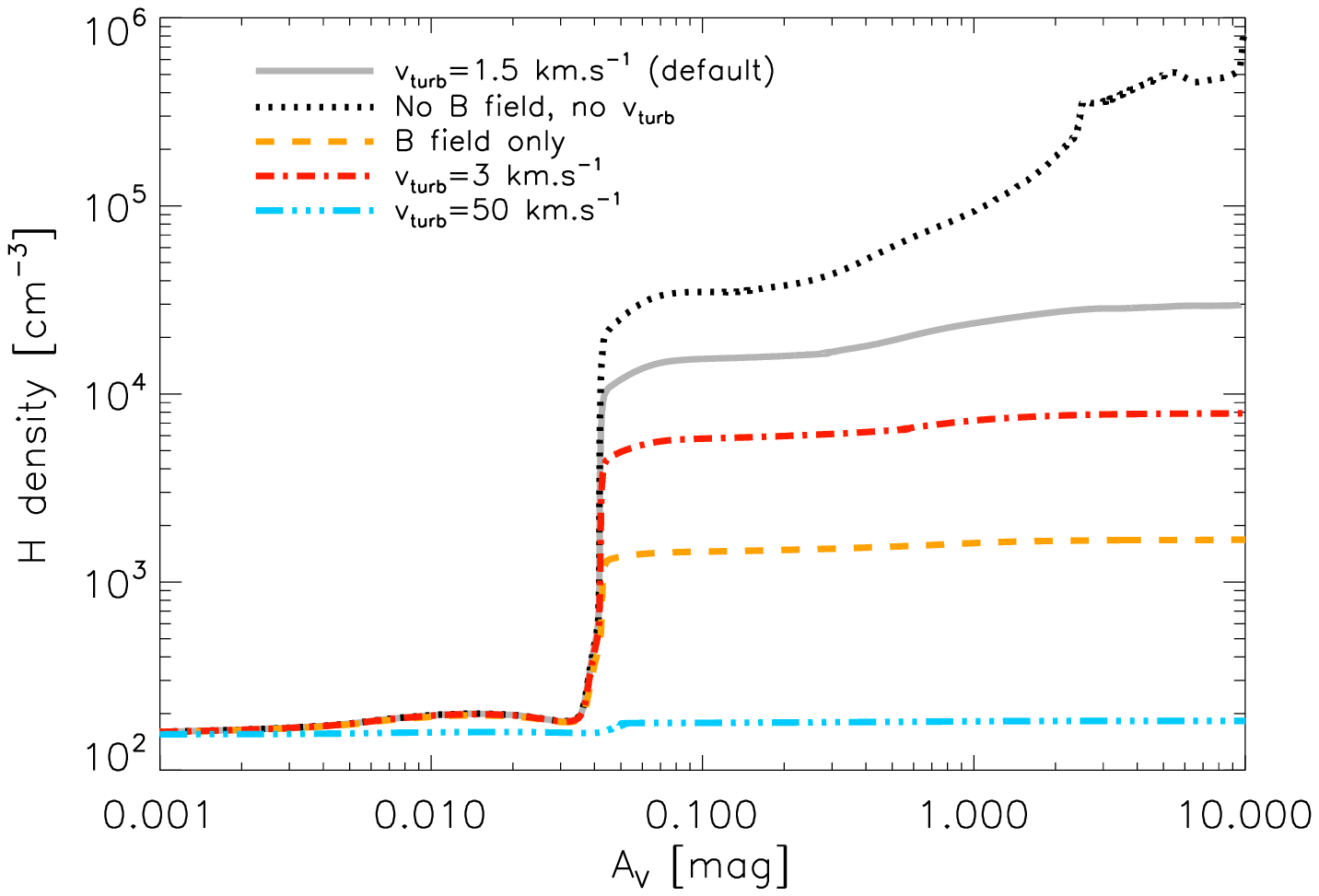} \hspace{1cm}
\includegraphics[clip,width=8cm]{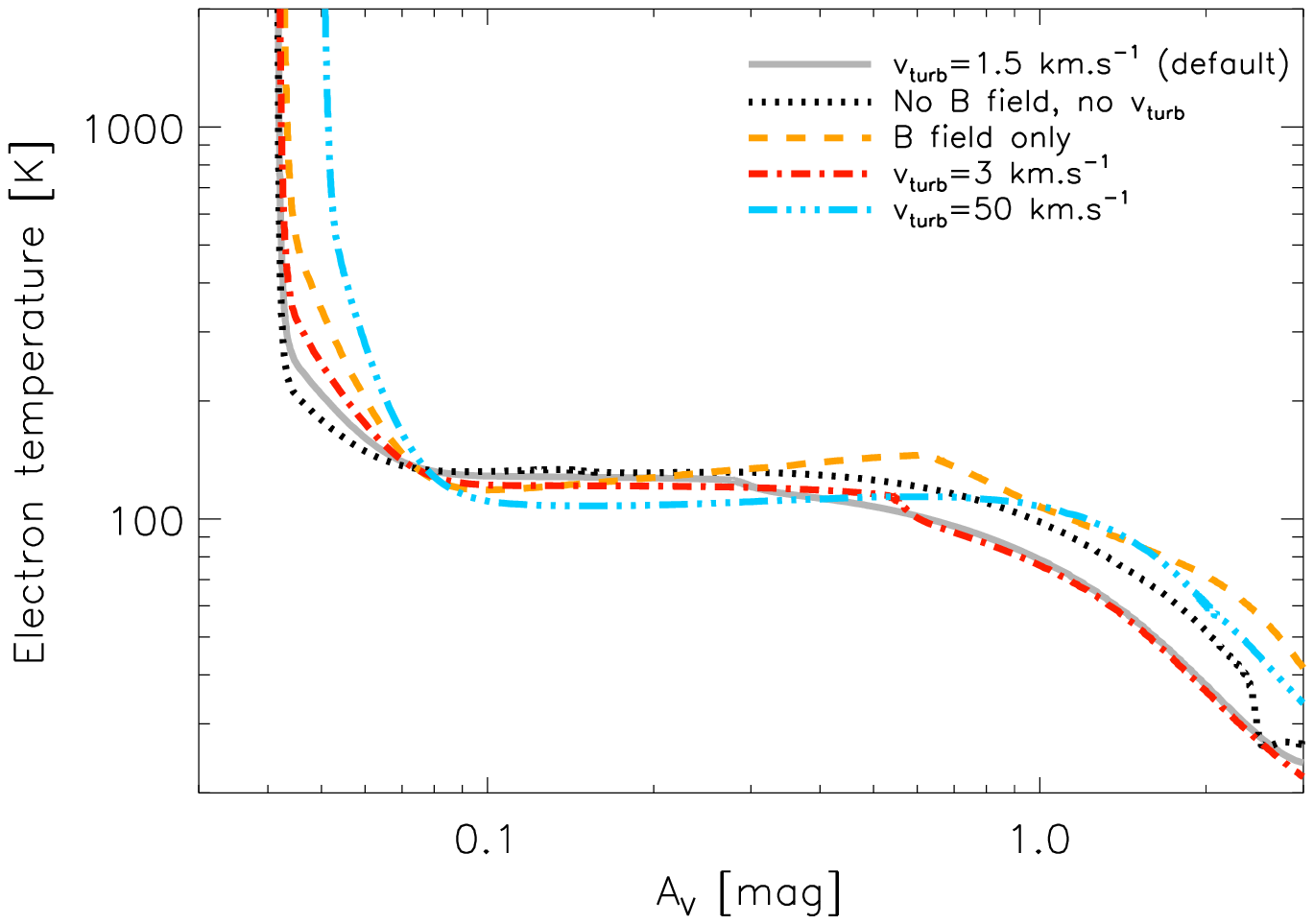}
\vspace{-5pt}
\caption{Effect of turbulence and magnetic fields on the predicted line intensities (top panel), density, and temperature (bottom panels) in the modeled cloud for the central region single-burst case. Green bars: observations. Gray bars and solid lines: only low turbulence switched on ($v_{\rm turb}$=1.5\,km\,s$^{-1}$, default model). Black bars and dotted lines: no magnetic fields and no turbulence. Orange bars and dashed lines: only magnetic field switched on (B=30\,$\mu$G). Red bars and dash-dotted line: only moderate turbulence switched on ($v_{\rm turb}$=3\,km\,s$^{-1}$). Cyan bars and triple-dotted-dashed line: only high turbulence switched on ($v_{\rm turb}$=50\,km\,s$^{-1}$).}
\label{turb}
\end{figure*}

The effects of magnetic field and turbulence on the cloud density, temperature, and line emission are shown in Fig.\,\ref{turb}. 
We present a set of runs for our best-fitting model in the central region single-burst case (note that we recover similar behaviors for the central continuous and southern single-burst cases, as shown in Appendix~\ref{appendixa}) with the magnetic field (B=30\,$\mu$G) and/or the turbulence pressure ($v_{\rm turb}$=1.5, 3, 50\,km\,s$^{-1}$) terms switched on. These terms have no impact on the ionic line emission because thermal pressure dominates the pressure balance in the \hii region, but they noticeably change the emission of the PDR lines. When only thermal pressure is considered, the gas density jumps to values $>3\times10^4$\,cm$^{-3}$ in the PDR and the \oi lines are overpredicted by an order of magnitude (black bars and dotted line). 
With only the magnetic field on, all three PDR lines are underpredicted by a factor of $\sim$3 (orange bars), but their ratios are kept in the range observed thanks to the lower densities achieved ($2\times10^3$\,cm$^{-3}$). Comparing models with different turbulent velocities, we see that the \cii line is best matched for low
or intermediate velocities because of their moderate densities ($\sim10^4$\,cm$^{-3}$) and slightly lower PDR temperatures (at $A_{\rm V} \simeq 1-3$\,mag). Increasing the turbulent velocity reduces the predicted \oi emission and PDR density. The high-turbulence model performs poorly because it has the most dramatic effect on the density and line emission. 

To summarize, we find that the best case that simultaneously matches all three PDR lines in the central region is the model with intermediate turbulent velocity ($v_{\rm turb}$=3\,km\,s$^{-1}$), which has a density of $8\times10^3$\,cm$^{-3}$, but we stress that the main effect of turbulence is to reduce the PDR density.

\subsection{Input spectra and SED}
\label{insed}
\begin{figure*}
\centering
\includegraphics[clip,trim=0 0 0.1cm 0,width=6.7cm]{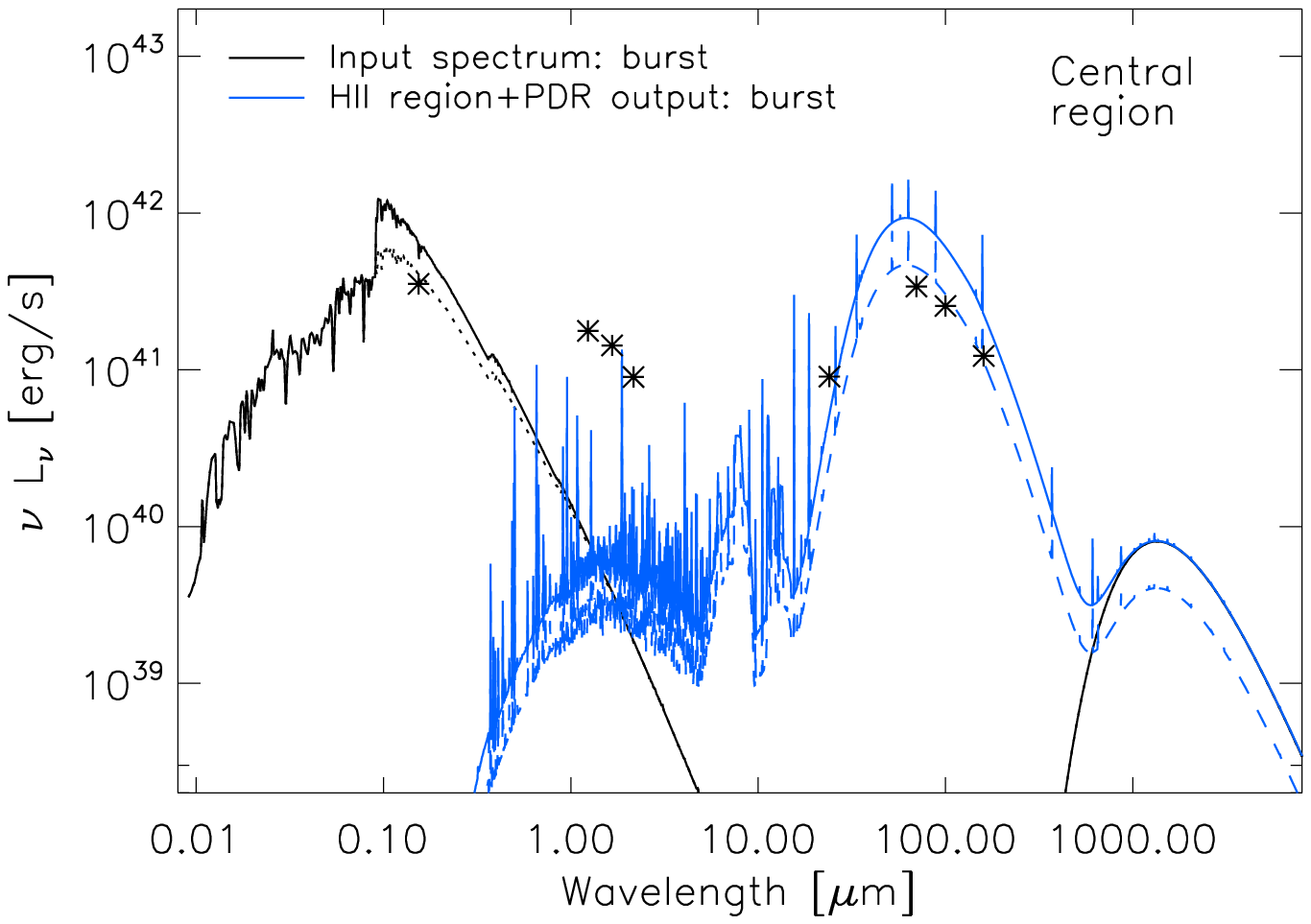}
\includegraphics[clip,trim=1.95cm 0 0.1cm 0,width=5.75cm]{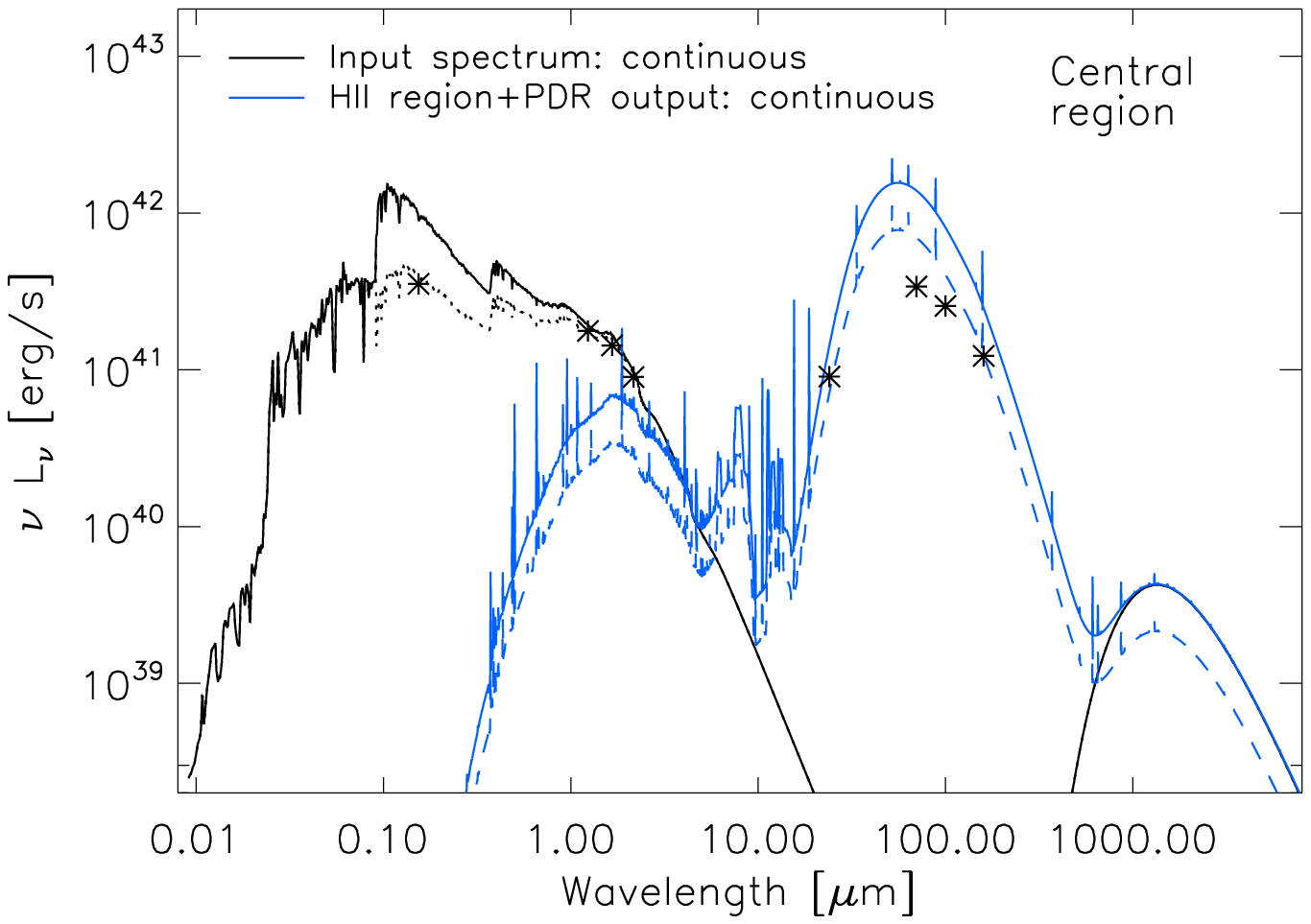}
\includegraphics[clip,trim=1.95cm 0 0.1cm 0,width=5.75cm]{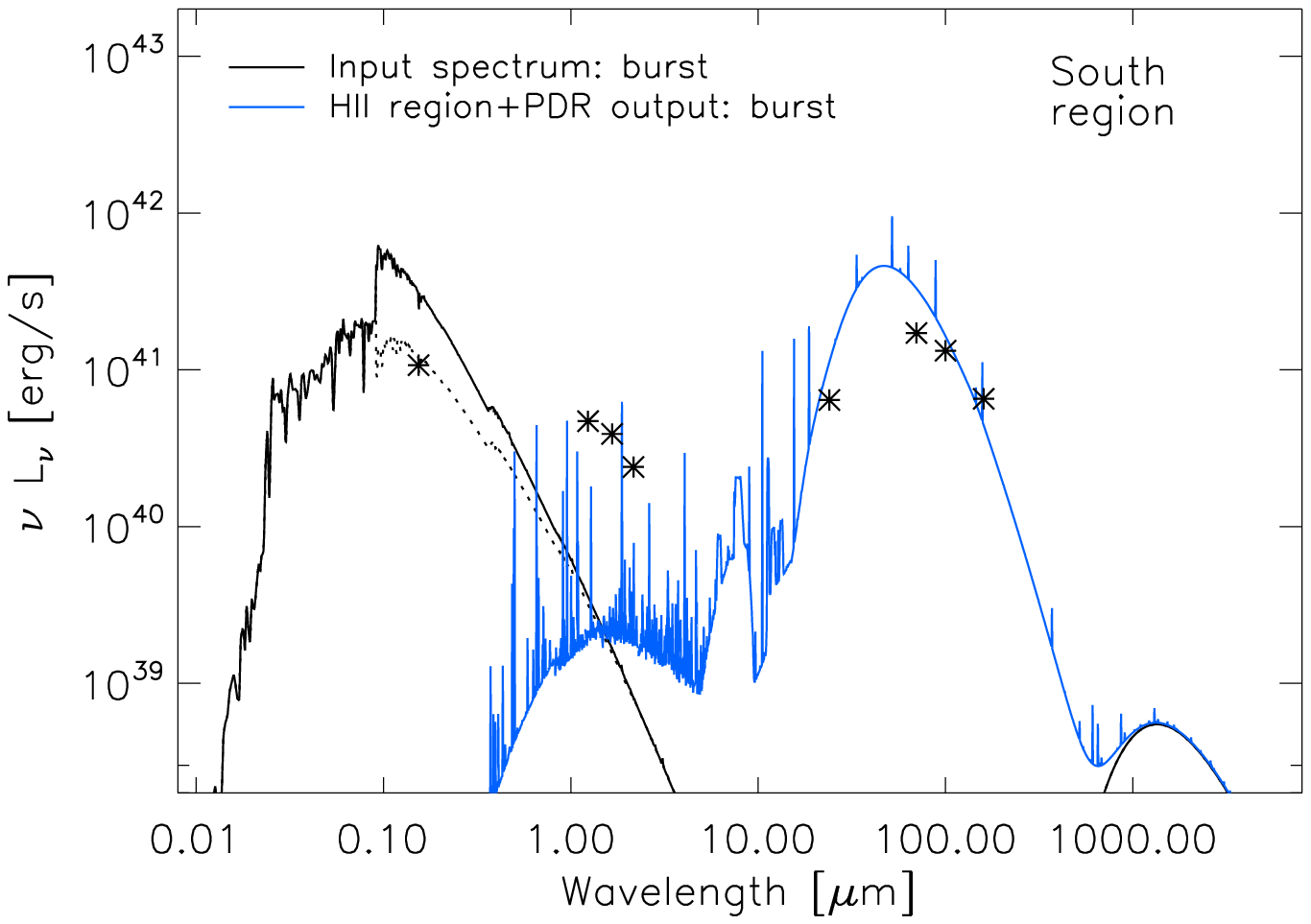}
\vspace{-10pt}
\caption{Spectral energy distributions of the two star-forming regions: central region single-burst case (left panel), central region continuous case (middle panel), and southern region single-burst model (right panel). The black and blue curves correspond to the input and output SEDs, respectively. The dotted lines are the attenuated input SEDs. The data points are the photometry measurements from GALEX FUV, 2MASS J, H, K bands, MIPS 24\,\mum, and PACS at 70\mum, 100\mum, and 160\,\mum. In panels for the central region, the dashed curves are scaled versions of the output SEDs, considering a covering factor of $0.5$ for the PDR. }
\label{spectra}
\end{figure*}

Figure~\ref{spectra} shows the input and output SEDs of the models for the two regions. The input SED is the stellar spectrum of the illuminating source modeled with \textsc{Starburst99} and also includes the CMB at millimeter wavelengths. In the central region, the stellar spectrum has a wider distribution for the continuous star formation model (red curve) than the single-burst star formation model (black curve), and it is more luminous in the near-IR regime due to the presence of old stars.  
Compared to observations, all input SEDs fall above the GALEX FUV data because they are unattenuated, and the single-burst input SEDs fall below the 2MASS data because they lack old stars. In the central region, the input SED of the continuous model, on the other hand, agrees well with the 2MASS data. For better agreement with the FUV data, we estimated the level of extinction required to attenuate the input SEDs. We considered average extinction values $E(B-V)$ of 0.1, 0.05, and 0.1\,mag for the central continuous central single-burst and southern single-burst models, respectively (dotted lines in Fig.\,\ref{spectra}), which are in the range of values found by \cite{ubeda-2007b}. 

Focusing on the output SEDs of the models for the central region, it is anticipated that the level of the FIR continuum is different. The higher the input luminosity of the source, the higher the peak of the output SED. Moreover, as the dust temperature rises, the peak of the output SED is expected to shift to shorter wavelengths. For the continuous model, the higher FUV luminosity therefore
provides more dust heating, explaining the slight shift of its peak to shorter wavelengths. 
We can compare the output SEDs to the observed PACS 70\mum, 100\mum, and 160\mum fluxes. The models agree relatively well with observations for the southern region, but overpredict the FIR continuum emission in the central region.
This is not surprising because we used as input luminosity a higher value than the observed TIR flux and the modeled PDR has a high $A_{\rm V}$. Better agreement with continuum observations requires a model that predicts a TIR flux lower by a factor of 2-3, for example, by reducing the covering factor of the PDR. We return to this in Sect.~\ref{modelum}.

\section{Discussion}
\label{discussion}
We have presented models for the two star-forming regions of NGC\,4214 that work for most of the observed MIR and FIR lines. Some discrepancies remain between our models and observations (\neii, \nii, \cii/\oi, and $L_{\rm TIR}$); these are not due to the choice of parameter space but rather to missing physics or components in our models. In this section, we discuss various aspects of our analysis: 1)~the discrepancies with observations, and we give clues to improve our models, 2)~which star formation scenario describes the data better, and 3)~how our ISM results relate to the known properties and evolution of the two regions.

\subsection{Discrepancies between models and observations}
\label{disclines}
\subsubsection{Ionic lines}
In our best-fitting models of the \hii region, the \neii12.8\mum and \nii122\mum lines are systematically underpredicted. These lines have lower excitation potentials (21.56 and 14.53\,eV, respectively) than the other, better matched ionic lines, and can therefore be partially excited outside of the main \hii region. This effect can be significant in NGC\,4214 because of the poor spatial resolution. 

Discrepancies between the observed and modeled \neiii/\neii ratio, with the same amplitude as we found for NGC\,4214, were reported by \cite{martin-hernandez-2002} for \hii regions observed by ISO/SWS (see their figure~2). These could again be related to a mixture of physical conditions within the ISO beam, which is also relatively large. For the starburst galaxy Haro\,11, we have examined the effect of an additional low-ionization component (star of effective temperature 35\,000\,K and $n_{\rm H}\simeq 10^{1-3}$\,cm$^{-3}$). This reproduced the observed \neii12.8\mum and \nii122\mum emission without significantly affecting the other ionic lines \citep{cormier-2012}.
Alternatively, given that the neon lines have the largest energy difference, they are more sensitive to the underlying stellar atmosphere models than the sulfur lines. Constraining those models is beyond the scope of this paper, and we relied on the sulfur lines as being more robust diagnostics of the \hii region conditions in NGC\,4214 (i.e., models presented in Sect.~\ref{modelresults}). 

We also assessed the effect of including the \neii12.8\mum line in the best-fitting solution-tracking procedure for the southern region. This resulted in a solution where the \neii12.8\mum absolute flux was better reproduced (within a factor of 3). This model uses an older burst (5\,Myr), higher density (500\,cm$^{-3}$), and the same inner radius (20\,pc). The corresponding ionization parameter is $\log(U)=-2.4$, which is 0.1\,dex lower than previously found. However, the model underpredicts the neon intensities by a factor of 2 and the [\textsc{S\,iv}]/[\textsc{S\,iii}]\,18.7\mum ratio by a factor of $\sim$4, while that ratio was matched within 20\% without the \neii constraint.

\subsubsection{PDR lines}
In our default PDR solutions for both regions, \cii is systematically underpredicted compared to \oi. The best PDR model, found in Sect.~\ref{magn} for the central region single-burst case, includes a turbulent velocity of 3\,km\,s$^{-1}$. In the southern and central region continuous case, similar turbulent velocities also lead to better agreement with the \cii/\oi line ratios, but still underpredict the observed emission in absolute values (Appendix~\ref{appendixa}).

In addition to the stars, X-rays can be a source of heating in the PDR and affect the FIR line emission. Point sources and diffuse X-ray emission have been reported in \cite{hartwell-2004} and \cite{ghosh-2006}. The identified point sources are not coincident with the peak of the FIR emission, and we therefore ignored them. The diffuse emission is mostly detected in the central region, with a luminosity of $3\times 10^{38}$\,erg\,s$^{-1}$ \citep{hartwell-2004}, which is lower than that of the starburst. We have tested the effect of this diffuse X-ray component on the PDR lines in the central region and found that it increases the predicted intensity of the \oi lines by $\sim$30\% and the \cii intensity by less than 10\%. As X-rays are not the main source of heating in the PDR, they do not help to produce significantly more \cii emission. 

We further explored the possibility of \cii originating from a diffuse ionized component. We compared the \cii and \nii122\mum intensities and the PACS upper limit on the \nii205\mum line, which gives \nii122\mum/\nii205\mum$>$1, to theoretical predictions assuming pure collisional regime. Following \cite{bernard-salas-2012} and applying $C$ and $N$ elemental abundances observed in NGC\,4214, we found that less than 16\% of the total observed \cii emission arises in diffuse ionized gas. \hers SPIRE FTS observations of \nii205\mum toward the central region also indicate that \nii122\mum/\nii205\mum$=$2.5 (priv. comm. R. Wu), which gives an ionized gas density of $\sim$60\,cm$^{-3}$ and a contribution of only 8\% to the total \cii emission. Therefore, if a low-density ionized component is added to our current models to account for the missing \nii and \neii emission, this component will not contribute significantly to the \cii emission. Our best-fitting models for the central region also predict that the \hii region contributes less than a few percent to the \cii emission. 

\begin{figure*}
\centering
\includegraphics[clip,trim=0 10mm 0 0,width=\textwidth,height=5.8cm]{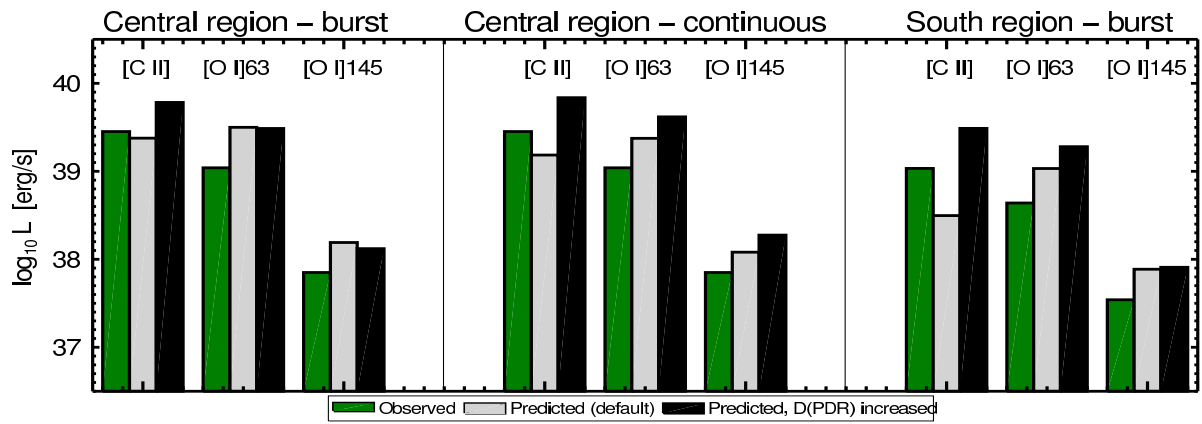}
\vspace{-10pt}
\caption{Effect of changing the distance to the modeled cloud for the PDR calculation. The PDR distance is increased by a factor of 2, 3, and 5 for the central single-burst (left panel), central continuous (middle panel), and southern single-burst cases (right panel), respectively.}
\label{pdrdist}
\end{figure*}

The \cii emission most likely arises from a neutral phase, but its conditions are not well described by our default PDR models. With the PDR tests performed in Sect.~\ref{magn}, we have explored the effect of density on the PDR emission lines, but these lines are also sensitive to the radiation field strength. To reduce the radiation field intensity $G_0$ (not the hardness), we placed the PDR farther away by stopping the model at the H$^+$/\hi phase transition and resuming the calculation at a larger distance in the \hi phase (note that this breaks the pressure equilibrium). The direct effect is to dilute the UV field before it reaches the PDR, as proposed by \cite{israel-2011} and \cite{cormier-2015}, which is equivalent to increasing the porosity of the medium. For the central single-burst central continuous and the southern single-burst cases, we increased the PDR distance by a factor of 2, 3, and 5 ($r_{\rm in}$$\simeq$170, 186, and 115\,pc), respectively. This way, $G_0$ decreases to $\sim$120 in all three cases, boosting the \cii emission, and the predicted \cii/\oi ratios match the observed ratios within 40\% (Fig.~\ref{pdrdist}). In absolute values, the \cii and \oi intensities are 2 to 3 times too high,
however. This can be compensated by a PDR covering factor lower than unity (see Sect.~\ref{modelum} below). 

We conclude that a moderate density, moderate $G_0$ neutral medium (compared to our dense, high $G_0$ default PDR model) with a
low turbulent velocity and a covering factor lower than unity is the most plausible origin for the observed \cii emission. 

\subsubsection{Model luminosity}
\label{modelum}
\cite{hermelo-2013} have fit the dust SED of the whole galaxy. In particular, their models require less UV luminosity than that observed to match the IR emission. They discussed possible explanations for this disagreement, proposing an escape of unattenuated UV radiation along with a particular geometry and dust properties in the galaxy. Such an argument of a UV escape fraction could also apply in our case. As seen in Fig.\,\ref{spectra}, the observed photometry data points in the FIR, which correspond to emission originating from the PDR, are lower than the modeled SED for the central region. This disagreement can indicate a different covering factor for the PDR. In our model, the PDR fully covers the sphere around the source (i.e., covering factor of unity). For the best-fitting models to better match the photometry, it could be that the PDR component is more porous, allowing radiation to escape the cloud. To illustrate this, we plot the resulting SEDs for the two models of the central region considering a PDR covering factor of 0.5 (dashed curves in Fig.\,\ref{spectra}).

Part of the discrepancies in our results that we have discussed in this section~\ref{disclines} originate in modeling each complex as a single cloud, which is imposed by the lack of spatial resolution in the observations. Clearly, future improvements are expected from observations with better spatial resolution.

\subsection{Central region: bursty or continuous star formation?}
\label{borc}
The star formation history of NGC\,4214 over the last Gyr is complex. It shows bursts lasting for shorter or longer periods and a continuous `background', which takes place throughout this whole time window \citep{mcquinn-2010,williams-2011}. In that sense, we could say that it is a rather hybrid star formation pattern. 
In the central region, we have investigated cases of both a single-burst and a continuous star formation mode. However, when we use a single model to reproduce the observed line emission, we simplify the problem, since we do not take into account both modes. Hence arises the question of which of the two approaches is the most adequate to model the MIR-FIR line emission. 
We have shown that both modes can satisfyingly reproduce the observed mid- and far-infrared line emission. By comparing the $\bar\chi^2$ values of the best-fitting models (Table~\ref{chi2ionic}), we found that the single-burst case seems to globally
perform better when considering the lines used in the optimization method. When including \nii and \neii, the two modes perform similarly, although the high $\bar\chi^2$ values are driven by the poor fit to the \neii12.8\mum line. 
For the PDR, the continuous scenario gives a lower $\bar\chi^2$ , but the default PDR solutions are not optimum and can be fine-tuned for both modes (by lowering the density and $G_0$, see Sects.~\ref{magn} and \ref{disclines}). We conclude that both are limiting, simplifying cases of modeling the ISM in NGC\,4214, with the continuous star formation model being marginally more accurate inside the PDR and the single-burst in the \hii region (without further refinement).

\subsection{Comparison between the two star-forming regions}
How do the ISM conditions that we have characterized in the two star-forming regions relate to their star formation properties? 
We have found that the modeled cluster (radiation source) in the southern region contains younger stellar populations with a harder radiation field than that in the central region, in agreement with the results of \cite{ubeda-2007}, for instance. The hydrogen density is also higher in the southern star-forming region, but the metallicities of the regions are very similar. The southern region is observed to be at a younger, more compact stage than the central region. The central region is more evolved
and had time to expand, as observed by the presence of shells that may have swept away the dense material \citep{walter-2001}, and is thus consistent with a more diffuse ISM. 

We calculated the star formation rate surface densities for the two regions combining the GALEX FUV map and the \spit 24\mum map, as done in \cite{leroy-2008}, with%
\begin{equation}
\label{sfcomp}
\begin{split}
\Sigma_{\rm SFR} {\rm [M_{\odot}\,yr^{-1}\,kpc^{-2}]} = 3.2 \cdot 10^{-3} \cdot {\rm I_{24}~[MJy\,sr^{-1}]} \\ ~+~ 8.1 \cdot 10^{-2} \cdot {\rm I_{\rm FUV}~[MJy\,sr^{-1}]}
\end{split}
,\end{equation}
\noindent where $\Sigma_{\rm SFR}$ is the SFR surface density and I$_{24}$ and I$_{\rm FUV}$ the 24\mum and FUV intensities. We also measured the atomic and molecular hydrogen content of the two regions using the 21cm map from THINGS\footnote{\url{http://www.mpia-hd.mpg.de/THINGS/Data.html}} \citep{walter-2008} and the CO(1-0), CO(2-1) transition maps from \cite{walter-2001} and HERACLES\footnote{\url{http://www.cv.nrao.edu/~aleroy/heracles_data/}} \citep{leroy-2009}, respectively. We used a conversion factor of $\alpha_{\rm CO}$=4.38~[M$_\odot$\,pc$^{-1}$\,(K\,km\,s$^{-1}$)$^{-1}$] from CO(1-0) luminosity to H$_2$ mass. If we were to use a different conversion factor due to the low metallicity of these regions \citep[e.g.,][]{schruba-2012}, this would not affect the relative comparison of the regions (see Table~\ref{sfprops}). 
The central region (I) has a total hydrogen content of M$_{\rm gas,I}$=M$_{\rm HI}$+M$_{\rm H_2}$=2.05$\times$10$^6\,{\rm M_\odot}$ and a molecular (mass) fraction of $f_{\rm mol,I}$=M$_{\rm H_2}$/M$_{\rm HI}$=0.35. Integrating Eq.~\ref{sfcomp} in the region, we find SFR$_{\rm I}$=2.2$\times$10$^{-2}\,{\rm M_\odot\,yr^{-1}}$. The southern region (II) has a higher total hydrogen content of M$_{\rm gas,II}$=2.68$\times$10$^6\,{\rm M_\odot}$, an H$_2$ fraction $f_{\rm mol,II}$=0.32, and SFR$_{\rm II}$=1.9$\times$10$^{-2}\,{\rm M_\odot\,yr^{-1}}$. 
Therefore the southern region has relatively more gas compared to its SFR than the central region. In terms of efficiency, SFR/M$_{\rm gas}$, it is about 50\% lower in the southern region. 
This could reflect a slightly more efficient, cluster-like star formation episode in the central region or simply encode a different evolutionary state as the SFR and gas masses are a strong function of time on scales of individual star-forming regions \citep[e.g.,][]{schruba-2010}. The southern region being younger, it may still be in the process of forming stars.

\begin{table}
  \caption{Comparison of star formation properties.}
\label{sfprops}
 \centering
\vspace{-5pt}
   \begin{tabular}{l c c c}
   \hline 
    \vspace{-10pt}\\
        Quantity                        & & Region I    & Region II   \\
        \hline
    \vspace{-10pt}\\
        {$L_{\rm CII}/L_{\rm TIR}$}     & & $5.4\times10^{-3}$  & $3.6\times10^{-3}$ \\
        {$L_{\rm CII}/L_{\rm CO(1-0)}$}         & & $6.7\times10^4$     & $2.5\times10^4$ \\
        {M$_{\rm gas=H_2+HI}$~$[$M$_\odot]$}    & & $2.05\times10^6$    & $2.68\times10^6$ \\
        {M$_{\rm H_2}$/M$_{\rm HI}$}    & & 0.35        & 0.32 \\
        {SFR~$[$M$_\odot$\,yr$^{-1}]$}  & & $2.2\times10^{-2}$  & $1.9\times10^{-2}$ \\
        {SFE~$[$Gyr$^{-1}]$}    & & 10.7        & 7.1 \\
        \hline 
   \end{tabular}
 \end{table}

The main differences in ISM conditions that we extracted from our modeling relate to the \hii region properties. The emission lines are a factor of two lower in luminosity in the southern region, except \neiii and \siv, which are proportionally higher in the southern region. By contrast, the PDR conditions in the two regions are similar ($n_{\rm H}\simeq10^4$\,cm$^{-3}$ and $G_0\simeq150$). Our modeling reflects conditions resulting from the recent star-forming event and has little predictive power regarding a different, future star-forming event. In particular, at the linear scale that we probe ($\sim$175\,pc), the PDR conditions are averaged over multiple star-forming clouds and not representative of the underlying, possibly different, substructure in individual molecular clouds. 
However, there is more PDR emission relative to CO in the central region, that is, high \cii/CO and \oi/CO ratios \citep[see also][]{cormier-2010}. 
As found by \cite{walter-2001}, CO emission is centrally concentrated in the south and more diffuse in the center. The concentration of molecular gas may be nourishing the current star formation episode in the south or is being observed at a pre-disruption stage with the same fate as the central region. 
The increased porosity, which evidently is an intrinsic property of the low-metallicity ISM, is seen in both the central and southern star-forming regions. 
The main evolution within the dense medium is seen in the covering factor of the PDR, which is found to be lower in the central, more evolved region than in the southern region, and in CO, which probably suffers more from photodissociation with time and its emission is seen farther away from the cluster center, but this cannot be modeled with our static approach. Observing an intermediate PDR tracer at the C$^+$/CO transition, such as C\,{\sc i}, would help to test this evolution scenario.

\section{Conclusion}
We have investigated the physical conditions characterizing 
the ISM of the dwarf irregular galaxy NGC\,4214 by modeling \spit 
and \hers observations of MIR and FIR fine-structure cooling lines. 
We used the spectral synthesis code \textsc{Cloudy} to 
self-consistently model the \hii region and PDR properties of the two main 
star-forming regions in NGC\,4214. 
We summarize our results as follows: 
\begin{itemize}
\item The ionized gas in the southern region is found to be $2.5$ 
times denser than in the central region (440\,cm$^{-3}$ 
versus 170\,cm$^{-3}$) and typified by a harder radiation field. 
Our best-fitting models of the \hii region+PDR reproduce most 
ionic and neutral atomic lines, namely the \oiii88\mum, \siii18.7 
and 33.5\mum, \siv10.5\mum, \neiii15.6\mum, and \oi63\mum lines, 
within a factor of $\sim$2. 
\item The observed \nii122\mum and \neii12.8\mum lines are 
the most discrepant with our model solutions for the \hii region. 
A single model component seems too simplistic to account for all 
observed lines simultaneously. Given the complexity of these star-forming regions, 
a multi-component modeling would be more appropriate. 
In particular, a lower excitation ionized gas component may be 
required to match the \nii and \neii emission in both regions. 
\item Our \hii region models and the established observational 
\cii/\nii line ratio used as a proxy for the fraction of \cii arising in the ionized gas 
both indicate that the \cii emission is mostly associated with the PDR gas, 
with only a $\sim$10\% contribution from the ionized gas. 
\item Constant pressure models where thermal pressure dominates 
the pressure equilibrium perform rather poorly for the PDR lines 
mostly because of the high densities and high $G_0$ values 
reached in the PDR. Including additional pressure terms, such as 
weak turbulent or magnetic pressure, or placing the 
PDR cloud farther away and reducing its covering factor, 
leads to a much improved reproduction of the observed line intensities. 
\item Star formation histories have an effect on the predicted 
MIR-FIR line emission. We have explored the two simplifying cases 
of a bursty and a continuous star formation scenario. 
In the central region, we found that the bursty scenario works marginally 
better for the \hii region and the continuous scenario for the PDR, 
although both modes can reproduce 
the observations after refining the PDR conditions. 
\item The \hii region modeling from IR emission is consistent with 
the evolutionary stages of the regions found in previous studies: 
the southern region is younger and more compact, while the central region 
is more evolved and diffuse. 
On the linear scale of our study ($\sim$175\,pc), the PDR conditions 
of individual star-forming clouds are averaged out and do not echo 
the observed differences between the two regions (stellar ages, \hii conditions, etc.). 
The increased porosity of the star-forming regions appears 
as an intrinsic characteristic of the low-metallicity ISM, 
while the covering factor of the PDR, which is reduced 
in the central region, stands out as the main evolution tracer.
\end{itemize}
\vspace{3mm}

\begin{acknowledgements}
We would like to thank S. Hony for his help with the IRS data and 
for fruitful discussion, F. Walter for providing us the CO(1-0) data, 
and S. Glover for his advice on turbulence issues.
We thank the referee for his or her comments on the manuscript.
DC and FB acknowledge support from DFG grant BI 1546/1-1.
This work is based in part on archival data obtained with the 
Spitzer Space Telescope, which is operated by the Jet Propulsion Laboratory, 
California Institute of Technology under a contract with NASA.
\hers is an ESA space observatory with science instruments 
provided by European-led Principal Investigator consortia and 
with important participation from NASA.
\end{acknowledgements}

\bibliographystyle{aa}
\bibliography{aa26447}

\appendix
\section{Comparison of PDR model predictions and observations}
\label{appendixa}
We show in this appendix the effect of turbulence and magnetic fields on the predicted PDR line intensities for the central region continuous case and for the southern region single-burst case. Tests with turbulence and magnetic fields are presented in Sect.~\ref{magn} only for the central region single-burst case. 

\begin{figure}[!h]
\centering
\includegraphics[clip,width=8cm]{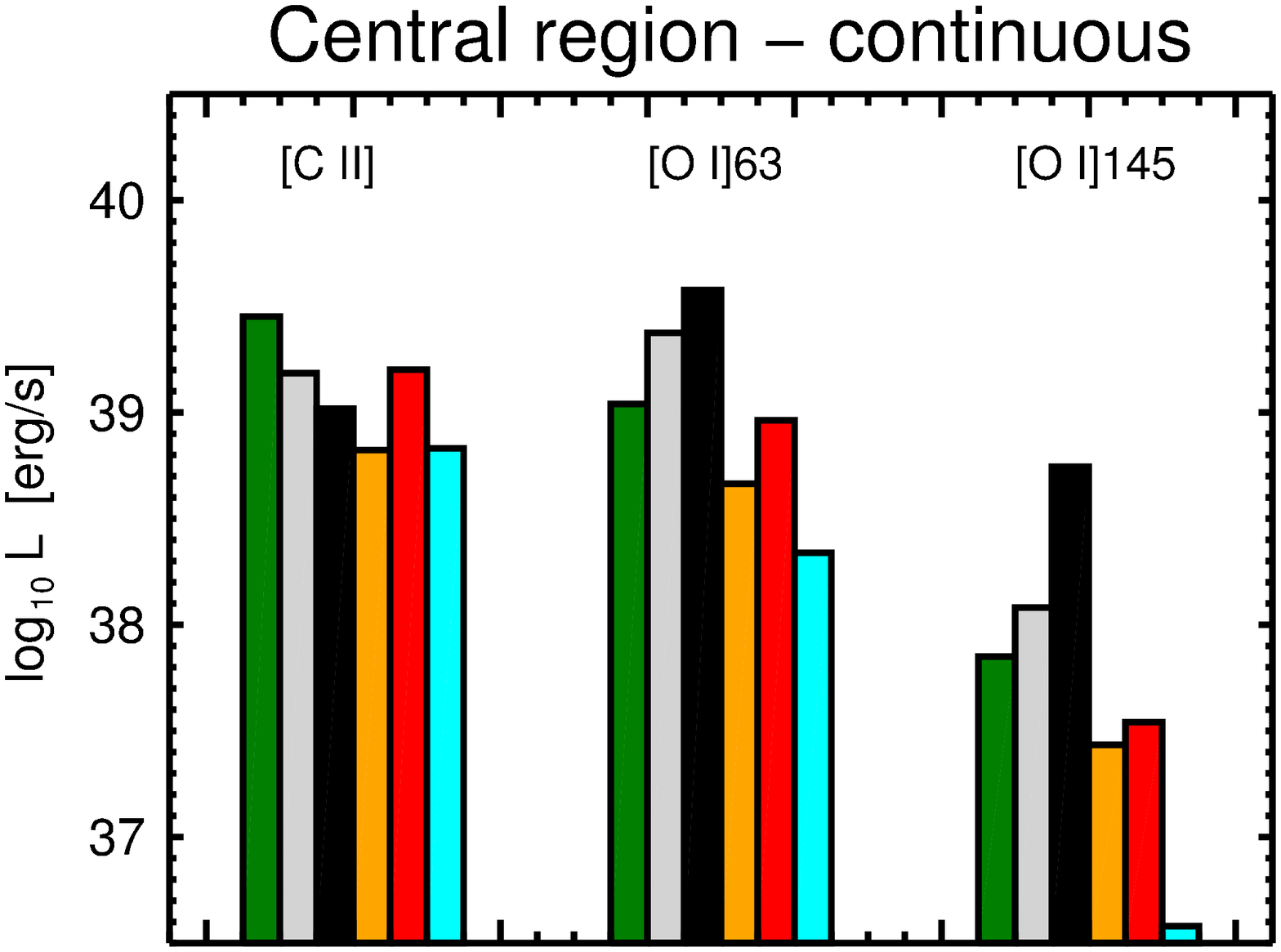}
\includegraphics[clip,width=8cm]{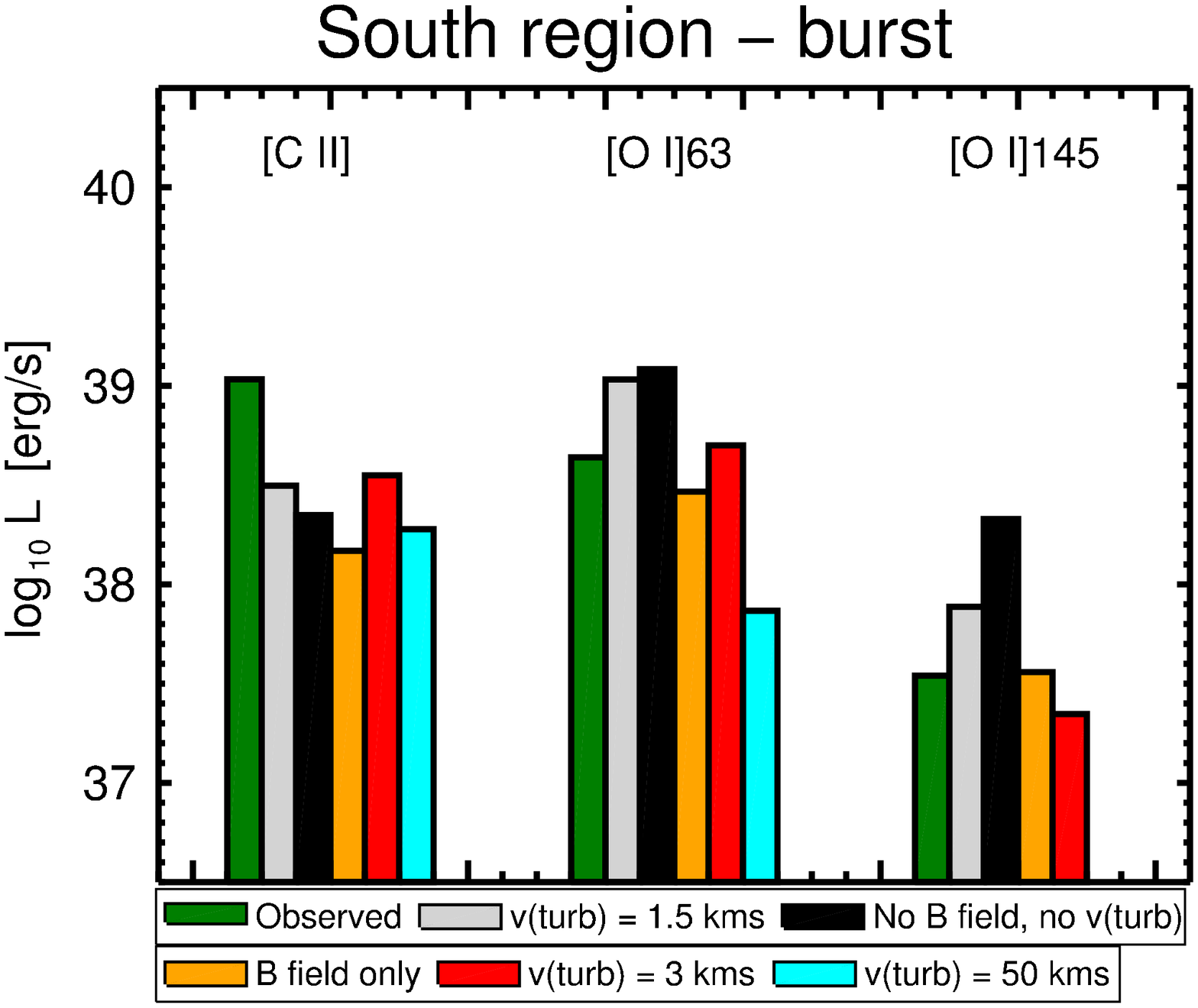}
\caption{
Same as Fig.~\ref{turb} for the central region continuous case (top panel) and for the southern region single-burst (bottom panel). See Sect.~\ref{magn} for details. 
}
\end{figure}

\end{document}